\documentclass[preprint, review, 10pt]{elsarticle}

\usepackage{graphicx}
\usepackage{fullpage}
\usepackage{mathrsfs}
\usepackage{amssymb}
\usepackage[normalem]{ulem}
\usepackage[dvipsnames]{xcolor}
\usepackage{comment}
\definecolor{mypink}{rgb}{0.858, 0.188, 0.478}

\usepackage{lineno}
\usepackage{verbatim}
\usepackage{xcolor}
\usepackage{eqnarray}
\usepackage{siunitx}
\usepackage[T1]{fontenc}
\usepackage{mathtools}
\RequirePackage{ifthen}
\RequirePackage{graphicx}
\RequirePackage{amsmath}
\RequirePackage{fancyhdr}
\usepackage{booktabs}
\usepackage{siunitx}
\usepackage{fancyhdr,graphicx,amsmath,amssymb,microtype}
\usepackage[ruled,vlined]{algorithm2e}
\SetKwRepeat{Do}{do}{while}
\usepackage{alltt}%
\SetAlFnt{\small}
\SetAlCapFnt{\normalsize}
\SetAlCapNameFnt{\normalsize}
\usepackage{algorithmic}
\algsetup{linenosize=\small}
\include{pythonlisting}
\usepackage{setspace}

\usepackage{booktabs}
\usepackage{multirow}

\usepackage{glossaries}
\newacronym{bc}{BC}{boundary condition}

\usepackage[utf8]{inputenc}
\usepackage[english]{babel}

\definecolor{purp}{rgb}{0.4,0.2,0.8}

\usepackage{caption}
\usepackage{subcaption}
\usepackage{hyperref}
\definecolor{custom-blue}{RGB}{3,69,173}
\hypersetup{
    colorlinks = true,
    urlcolor   = blue,
    citecolor  = custom-blue,
}

\journal{International Journal for Uncertainty Quantification}

\begin{document}

\begin{frontmatter}


% \title{Non-intrusive polynomial chaos on diffusion manifolds for fast uncertainty quantification in complex stochastic systems}
\title{Manifold learning-based polynomial chaos expansions for high-dimensional surrogate models}

%%% Alternate Titles
% Surrogates for high-dimensional models: Polynomial Chaos Expansions on Subspace Manifolds

% Polynomial Chaos Expansions on Subspace Manifolds

% Fast UQ in high dimensions: A novel encoder-decoder framework based on polynomial chaos expansions

% A novel polynomial chaos-based encoder-decoder framework for uncertainty quantification in high-dimensional models

% Surrogate modeling in high dimensions with an encoder-decoder framework based on polynomial chaos expansions and manifold learning

% Surrogates for high-dimensional models with polynomial chaos expansions on latent spaces


\author[1]{Katiana Kontolati}
\author[2,3]{Dimitrios Loukrezis}
\author[4]{Ketson R. M. dos Santos}
\author[1]{Dimitrios G. Giovanis}
\author[1]{Michael D. Shields\footnote{Corresponding Author}}

\address[1]{Department of Civil \& Systems Engineering, Johns Hopkins University, Baltimore MD, USA}

\address[2]{Institute for Accelerator Science and
Electromagnetic Fields, Technische Universität Darmstadt, Darmstadt, Germany}

\address[3]{Centre for Computational Engineering, Technische Universität Darmstadt, Darmstadt, Germany}

\address[4]{Earthquake Engineering and Structural Dynamics Laboratory, \'{E}cole Polytechnique F\'{e}d\'{e}rale de Lausanne, VD, Switzerland}

\begin{abstract}

In this work we introduce a manifold learning-based method for uncertainty quantification (UQ) in systems describing complex spatiotemporal processes. Our first objective is to identify the embedding of a set of high-dimensional data representing quantities of interest of the computational or analytical model. For this purpose, we employ Grassmannian diffusion maps, a two-step nonlinear dimension reduction technique which allows us to reduce the dimensionality of the data and identify meaningful geometric descriptions in a parsimonious and inexpensive manner. Polynomial chaos expansion is then used to construct a mapping between the stochastic input parameters and the diffusion coordinates of the reduced space. An adaptive clustering technique is proposed to identify an optimal number of clusters of points in the latent space. The similarity of points allows us to construct a number of geometric harmonic emulators which are finally utilized as a set of inexpensive pre-trained models to perform an inverse map of realizations of latent features to the ambient space and thus perform accurate out-of-sample predictions. Thus, the proposed method acts as an encoder-decoder system which is able to automatically handle very high-dimensional data while simultaneously operating successfully in the small-data regime. The method is demonstrated on two benchmark problems and on a system of advection-diffusion-reaction equations which model a first-order chemical reaction between two species. In all test cases, the proposed method is able to achieve highly accurate approximations which ultimately lead to the significant acceleration of UQ tasks. 

\end{abstract}


\end{frontmatter}

%%\begin{comment}

%\tableofcontents


\section{Introduction}
\label{S:Background}

% Review of Dimension Reduction for Surrogate Modeling Methodologies

Robust engineering design and optimal decision making require the accurate prediction of the performance of (often complex) stochastic systems or systems with significant uncertainty. Uncertainty quantification (UQ), an important field of computational science and engineering, provides a means of propagating uncertainties through the system to understand their influence on responses of interest. Despite the recent progress in hardware and processing power, UQ is often prohibitively expensive for real-world systems of interest, as it usually requires a large number of evaluations of complex mathematical models. To alleviate this issue, surrogate models are employed to establish an efficient approximate mapping between model inputs and outputs. Such models enable the propagation of mixed aleatoric and epistemic uncertainties across scales \cite{bhosekar2018advances}. The construction of accurate surrogates, however, typically requires smooth input-output functional relations, which may not be realistic in real-world applications, that predict low-dimensional quantities of interest that may not reflect the complexity of the solution.  

% Both limitations affect UQ tasks for large-scale models, which may additionally generate data that exhibit strongly nonlinear relations on complex, high-dimensional spaces (e.g., nonlinear, time-varying responses).

Non-intrusive polynomial chaos expansions (PCE) are an established and versatile surrogate modeling technique that express model input-output relations in terms of an expansion of polynomials that are orthonormal with respect to the probability density function (PDF) characterizing the input random variables \cite{bhosekar2018advances, xiu2002wiener, zhou2019expanded, hadigol2018least, babuvska2007stochastic}. One of the main advantages of PCE methods is that UQ tasks such as moment estimation and sensitivity analysis can be easily performed by post-processing the terms of the PCE \cite{sudret2008global, shao2017bayesian, crestaux2009polynomial}. In the case of a high-dimensional input parameter space, sparse PCE methods have been successfully proposed in the literature, which take advantage of the so-called ``sparsity of effects'' principle to construct surrogates with only a small number of forward model evaluations \cite{luthen2020sparse, blatman2011adaptive, alemazkoor2018preconditioning}. In such cases however, a suitable model selection criterion must be employed for tuning the hyperparameters that are used to obtain the optimal model. Several techniques exist to adaptively identify the optimal polynomial basis and associated sparse solution, which aim to keep the size of the basis small by controlling which functions are added to the basis \cite{jakeman2015enhancing, hampton2018basis, blatman2011adaptive, loukrezis2020robust}. Despite its advantages, in cases where very high-dimensional outputs are considered, a PCE surrogate can be both computationally intractable to construct and incapable of accurately performing out-of-sample predictions.

One way to overcome the challenges associated with high-dimensional models is to apply dimension reduction techniques. Linear and nonlinear dimension reduction methods can be used to map data onto lower-dimensional manifolds (embeddings) by identifying and extracting meaningful features. Such techniques are important to overcome the so-called ``curse of dimensionality'', to avoid overfitting, to denoise data, and to enable regression analysis tasks. Although dimension reduction methods were originally developed for computer vision and image recognition applications, they have been increasingly used in recent years to facilitate the construction of accurate surrogates for high-dimensional physics-based models. 

Several methods use linear spectral decomposition methods such as principal component analysis (PCA), which involve the eigendecomposition of the data covariance function to capture the dominant modes of the output represented in the form of a field (matrix) \cite{hombal2013surrogate, boukouvala2013reduced}. Similarly, proper orthogonal decomposition (POD) or the Karhunen-Loeve expansion (KLE) has been widely used for reduced-order model construction \cite{amsallem2008interpolation, zimmermann2013gradient, carlberg2011low, nath2017sensor}. Active subspaces, a dimension reduction technique which discovers linear manifolds of the data, has been proposed as an in-built technique for the construction of Gaussian process (GP) surrogates \cite{tripathy2016gaussian, vohra2020fast, constantine2014active, constantine2015active}. Furthermore, multiple gradient-based techniques can be found in the literature for identifying subspaces in situations involving multivariate outputs and high-dimensional input parameter spaces \cite{bigoni2021nonlinear, ji2018shared, zahm2020gradient, zimmermann2013gradient, carlberg2011low}. 

Nonlinear dimension reduction, also known as manifold learning is used to deal with the limitations of the linear methods; namely the assumption that high-dimensional data can be embedded in linear spaces. Instead, nonlinear dimension reduction methods consider that the data reside on some low-dimensional, nonlinear manifold such as a Grassmannian or a diffusion manifold. Recent work of the co-authors has considered the construction of surrogate models on the Grassmannian \cite{giovanis2018uncertainty, giovanis2020data, kontolati2021manifold}. Another class of methods leverages diffusion maps (DMaps) \cite{coifman2006diffusion}, to either draw samples from a distribution on the diffusion manifold \cite{ soize2016data, soize2017polynomial, soize2021probabilistic} or construct surrogate models on the diffusion manifold \cite{kalogeris2020diffusion, koronaki2020data}. Additionally, in a recent work kernel PCA is coupled with Kriging and PCE to extend surrogates to high-dimensional models \cite{lataniotis2020extending}. 

% Another approach involves the construction of GP surrogates from data projected on nonlinear manifolds representing orthogonality constraints \cite{giovanis2020data, kontolati2021manifold}. Diffusion maps (DMAP) , a method which uses sliding windows of statistic distributions of random walk increments to identify nonlinear embeddings, has been successfully employed for the construction of accurate emulators in a reduced space \cite{kalogeris2020diffusion, koronaki2020data}. 

An alternative approach to identify latent representations of data is to use deep neural networks (DNNs) such as multi-layer perceptrons (MLPs) \cite{schmidhuber2015deep, goodfellow2016deep}. Recently, multiple techniques based on autoencoders (unsupervised learning) \cite{wang2016auto} and convolutional neural networks (supervised learning) \cite{rawat2017deep}, have been proposed for constructing surrogate models when input and output fields are high-dimensional \cite{tripathy2018deep, nikolopoulos2021non, zhu2018bayesian, mo2019deep, thuerey2020deep, wang2021efficient, mo2019deep2, hesthaven2018non}. Such methods have lately received increasing attention primarily due to advancements in computer hardware and the availability of powerful resources such as graphical and tensor processing units (GPUs, TPUs). Although DNNs are capable of capturing complex nonlinear relations between high-dimensional inputs and outputs and provide both encoder and decoder paths for dimension reduction purposes, they are still considered more suitable for problems in the so-called ``big-data'' regime. Furthermore, such models are very costly to train, rely on the heuristic choice of the network architecture and the calibration of multiple hyperparameters. Finally, they do not inherently provide a link between the input stochastic parameters and model output, which is essential for the implementation of UQ tasks.  

In this work we introduce a novel framework, which combines low-dimensional manifold learning principles with surrogate model construction for the interpolation of dimension-reduced solutions that can be employed to generate out-of-sample predictions based on a limited number of model evaluations. We are interested in complex models that generate high-dimensional outputs (e.g., high-fidelity finite element models) that are computationally expensive to run. As a result, we can only afford a small number of model evaluations. Dimensionality reduction is achieved using the Grassmanian diffusion maps (GDMaps) technique introduced in \cite{dos2020grassmannian}, which identifies a latent representation of the dataset on a lower-dimensional manifold via a two-step procedure. In the first step, high-dimensional data (model solutions) are projected onto an orthonormal matrix manifold called the Grassmann manifold \cite{zhang2018grassmannian, edelman1998geometry} that defines the subspace structure of the data. In the second step the diffusion maps (DMaps) method is employed to unfold the underlying nonlinear geometry of the data on the Grassmann manifold onto a diffusion manifold. Next, a PCE surrogate model is constructed to establish a mapping between input parameters and coordinates on the diffusion manifold. To reconstruct the full solution from PCE prediction solutions on the diffusion, a set of special functions called geometric harmonics (GH) \cite{coifman2006geometric} are used to locally define suitable mappings from the data on the diffusion manifold onto the tangent space of the Grassmann manifold. The local GH models allow us to perform out-of-sample predictions and return generated points on the diffusion manifold to the physically interpretable space. 

%The advantages and contributions of the proposed method are listed as follows:
\begin{comment}
\begin{enumerate}
    \item Handles high-dimensional datasets by extracting important descriptors, which sufficiently represent the physics of the system;
    \item Adaptive PCE surrogate modeling technique allows us to identify a sparse representation on the diffusion manifold by detecting the significant PCE coefficients which results in a reduced computational cost as the number of input parameters increases;
    \item Enables the minimization of necessary model simulations as it works well in the small data regime;
    \item Provides a direct way to decode the compressed data to the original space and link input parameters with the model output;
    \item Greatly reduces the training time;
    \item Allows the acceleration of UQ tasks in cases of non-linear complex applications
\end{enumerate}
\end{comment}

The advantages of the proposed method lie in its ability to automatically handle high-dimensional datasets generated by complex models and extract important low-dimensional descriptors which sufficiently represent the complex physics of the system. Furthermore, our approach enables the minimization of necessary model simulations as it works well in the small-data regime, greatly reduces training time and provides a direct way to decode the compressed data to the original space and link input parameters with model outputs. We show that the proposed method is robust and allows the acceleration of UQ tasks in cases of non-linear complex applications.

The rest of this paper is organized as follows. The theoretical background for the dimension reduction methods, geometric harmonics, and PCE surrogates employed in this work, is briefly presented in Section \ref{prem}. The important ingredients of the proposed framework are discussed in detail in Section \ref{method}. This is divided into two sections where we describe an ``encoder path'' that follows the Grassmannian diffusion maps and PCE surrogates on the manifold, and a ``decoder path'' that describes the construction of local geometric harmonics to generate full solutions from reduced order predictions. The performance of the proposed approach is assessed by three illustrative applications given in Section \ref{examples}. The first example involves a model problem from electromagnetic field theory. In the second example, the method is applied to predict time-evolution on the classic Lotka-Volterra (predator-prey) dynamical system. The third application deals with a system of advection-diffusion-reaction equations modeling a first-order chemical reaction between two species. Finally, Section \ref{discussion} presents the conclusions.

The distinct components of the proposed method (GDMaps and PCE surrogate modeling) have been individually implemented in UQpy (Uncertainty Quantification with python) a general-purpose open-source software for modeling uncertainty in physical and mathematical systems \cite{olivier2020uqpy}. Codes for implementing the proposed framework and reproducing the results are available at: \url{https://github.com/katiana22/GDM-PCE}.

\section{Preliminaries}
\label{prem}

\subsection{Grassmannian Diffusion Maps (GDMaps)}

Diffusion maps (DMaps) \cite{coifman2006diffusion} is a manifold learning technique that is based on the construction of a Markov transition probability matrix corresponding to a random walk on a graph connecting the data. The vertices of the graph are the data points and the edges represent connections between the data points that are weighted by transition probabilities representing the local similarities between pairs of points. The graph structure can be parameterized by the so-called diffusion coordinates, representing the low-dimensional manifold (embedding) of the data. To identify this parameterization, a careful selection (ideally parsimonious \cite{dsilva2018parsimonious}) of the eigenvectors of the Markov matrix needs to be performed. 

Grassmannian diffusion maps (GDMaps) \cite{dos2020grassmannian} is a recently proposed variant of DMaps that defines similarity (or affinity) between very high-dimensional data points based on their underlying subspace structure and leverages DMaps to build a graph connecting subspaces on the Grassmann manifold \cite{absil2004riemannian, edelman1998geometry, ye2016schubert, ye2019optimization}. Herein, the basic elements of GDMaps are briefly presented and drawn from \cite{dos2020grassmannian, giovanis2018uncertainty}.

% circumvent limitations related to the accurate reveal of the underlying subspace structure. The method relies  on the construction of an affinity matrix between subspaces represented by points on the Grassmann manifold \cite{ye2016schubert, ye2019optimization}. Herein, the basic elements of GDM are briefly presented and drawn from \cite{dos2020grassmannian, giovanis2018uncertainty}.

\subsubsection{Grassmann manifold principles}

The \textit{Grassmann manifold} or \textit{Grassmannian}, denoted $\mathcal{G}(p,n)$ or $\mathcal{G}_{p,n}$, is the set of all $p$-dimensional subspaces embedded in $\mathbb{R}^{n}$.
% The orthogonal group $O(n)$ acts transitively on $\mathcal{G}(p,n)$ and as any $p$-dimensional subspace $\mathbb{W} \subseteq \mathbb{R}^{n}$ has an isotropy group isomorphic to $O(p) \times O(n-p)$. Thus, we obtain the characterization of the Grassmannian as
% \begin{equation}
% \label{eq:Grass}
%     \mathcal{G}(p,n) \cong \frac{O(n)}{O(p) \times O(n-p)},
% \end{equation}
% where `$\cong$' denotes a diffeomorphism, an invertible function that maps one differentiable manifold to another. 
$\mathcal{G}(p,n)$ is a smooth manifold of dimension $p(n-p)$. A point on the Grassmannian, $\mathcal{X} \in \mathcal{G}(p,n)$, is represented (the Stiefel representation) by an orthonormal matrix $\mathbf{X} \in \mathbb{R}^{n \times p}: \mathbf{X}^{T} \mathbf{X} = \mathbf{I}_{p}$ where $\mathbf{I}_{p} \in \mathbb{R}^{p \times p}$ is the identity matrix. That is, the point $\mathcal{X}$ is defined as the space spanned by the basis vectors $\mathbf{X}$, $\mathcal{X}=\text{span}(\mathbf{X})$.

For a group of points on the Grassmannian, the Riemmannian center of mass, also known as the \textit{Karcher mean} \cite{giovanis2020data}, is defined as the point $\mathcal{Y}$ that minimizes locally the cost function $\lambda :  \mathcal{G}(p,n) \rightarrow \mathbb{R}_{\ge 0}$ given by: 
\begin{equation}
    \lambda(\mathcal{Y})=\int_{\mathcal{G}_{p,n}}d_{\mathcal{G}_{p,n}}^2(\mathcal{Y},\mathcal{X})dP(\mathcal{X})
\end{equation}
where $dP(\mathcal{X})=\rho(\mathcal{X})d\mathcal{G}_{p,n}(\mathcal{X})$ is a probability measure over the infinitesimal volume element $d\mathcal{G}_{p,n}(\mathcal{X})$ with probability density $\rho(\mathcal{X})$ and $d_{\mathcal{G}_{p,n}}$ represents a distance measure on the Grassmannian. For a set of independent sample points $\{\mathbf{\mathcal{X}}_i\}_{i=1}^N \in \mathcal{G}(p,n)$, the sample Karcher mean $\mathbf{m}$ is estimated as the local minimizer of
\begin{equation}
    \lambda(\mathbf{m}) = \frac{1}{N} \sum_{i=1}^{N} d_{\mathcal{G}_{p,n}}^2 (\mathcal{X}_i, \mathbf{m}).
\label{eq:karcher1}
\end{equation}

Given the smoothness of the Grassmannian, one can define the \textit{tangent space} at a given point $\mathcal{X} \in \mathcal{G}(p,n)$, denoted $\mathcal{T}_{\mathcal{X}} \mathcal{G}(p,n)$, as the derivative of a trajectory $\gamma(z)$ on the manifold. The tangent space is represented by the set of all tangent vectors in $\mathcal{X}$, such that 
\begin{equation}
\label{eq:tangent}
    \mathcal{T}_{\mathcal{X}} \mathcal{G}(p,n) = \{ \mathbf{\Gamma} \in \mathbb{R}^{n \times p}: \mathbf{\Gamma}^\top \mathbf{X} = 0 \}.
\end{equation}
The trajectory $\gamma(z)$ is defined as the shortest (geodesic) path between two points, $\mathcal{X}_0$ and $\mathcal{X}_1$ on $\mathcal{G}(p,n)$. If $z$ is defined on the unit line, i.e., $z \in [0,1]$, the two points are denoted as $\gamma(0)=\mathcal{X}_0$ and $\gamma(1)=\mathcal{X}_1$. 

In the neighborhood of a point $\mathcal{X}_0$, mapping between the Grassmannian and the tangent space can be performed by the \textit{logarithmic} and \textit{exponential} mappings. 
% We define an exponential mapping $\text{exp}_{\mathcal{X}} : \mathcal{T}_{\mathcal{X}} \mathcal{G}(p,n) \rightarrow \mathcal{G}(p,n)$ to the tangent space for two points in $\mathcal{X} \in \mathcal{G}(p,n)$. 
% Consider a tangent space $\mathcal{T}_{\mathcal{X}} \mathcal{G}(p,n)$ with $\mathbf{\Gamma} \in \mathcal{T}_{\mathcal{X}} \mathcal{G}(p,n)$ and 
Consider two points $\mathcal{X}_0, \mathcal{X}_1$ on $\mathcal{G}(p,n)$ with $\gamma(0)=\mathcal{X}_0$, $\gamma(1)=\mathcal{X}_1$ represented by the matrices $\mathbf{X}_0$, $\mathbf{X}_1$, and $\dot{\gamma}(0) = \dot{\mathcal{X}_0}$ represented by the matrix $\mathbf{\Gamma}_0 \in \mathcal{T}_{\mathcal{X}_0} \mathcal{G}(p,n)$. One can map from a point $\mathcal{X}_1$ to the tangent space $\mathcal{T}_{\mathcal{X}_0} \mathcal{G}(p,n)$ through the logarithmic mapping
\begin{equation}
\label{eq:log}
    \text{log}_{\mathcal{X}_0}(\mathbf{X}_1) = \mathbf{\Gamma_1}=\mathbf{U}\tan^{-1}(\mathbf{\Sigma})\mathbf{V}^\top.
\end{equation}
where $\mathbf{\Gamma_1}\in \mathcal{T}_{\mathcal{X}_0} \mathcal{G}(p,n)$ and $\mathbf{U}$, $\mathbf{\Sigma}$, $\mathbf{V}$ are obtained from the singular value decomposition (SVD) of the matrix $\mathbf{M}=(\mathbf{X}_1-\mathbf{X}_0\mathbf{X}_0^\top\mathbf{X}_1)(\mathbf{X}_0^\top\mathbf{X}_1)^{-1}=\mathbf{U\Sigma V}$. Moreover, one can map from the point $\mathbf{\Gamma_1}\in \mathcal{T}_{\mathcal{X}_0} \mathcal{G}(p,n)$ to the point $\mathbf{X_1}$ through the exponential mapping: 
\begin{equation}
\label{eq:exp}
    \text{exp}_{\mathcal{X}_0}(\mathbf{\Gamma}_1) = \mathbf{X}_1=\mathbf{X}_0\mathbf{V}\cos(\mathbf{\Sigma}) + \mathbf{U}\sin(\mathbf{\Sigma}),
\end{equation}
where matrices $\mathbf{U}$, $\mathbf{\Sigma}$, $\mathbf{V}$ are defined by the SVD of $\mathbf{\Gamma}_1=\mathbf{U\Sigma V^\top}$. Additional details can be found in \cite{begelfor2006affine}.
% , and $\mathbf{Q}$ satisfies $\mathbf{V}\cos(\mathbf{\Sigma})\mathbf{Q}^\top = \mathbf{X}_0^\top \mathbf{X}_1$ and $\mathbf{U}\sin(\mathbf{\Sigma})\mathbf{Q}^\top = \mathbf{X}_1-\mathbf{X}_0\mathbf{X}_0^\top \mathbf{X}_1$ 

% One can map $\mathbf{\Gamma}$ to $\gamma(1)=\mathcal{X}_1$, where $\gamma(0)=\mathcal{X}_0$, $\dot{\gamma}(0) = \dot{\mathcal{X}_0}$, and $\dot{\mathcal{X}_0} \in \mathcal{T}_{\mathcal{X}_0} \mathcal{G}(p,n)$ is a vector defined on the manifold. The exponential mapping from the tangent space to the Grassmannian is expressed as
% \begin{equation}
% \label{eq:exp}
%     \text{exp}_{\mathcal{X}_0}(\mathbf{\Gamma}) = \mathbf{X}_1,
% \end{equation}
% where $\mathbf{X}_1$ is an orthonormal matrix which corresponds to point $\mathcal{X}_1$. Similarly we aim to define an explicit mapping from the Grassmannian to the tangent space. The inverse logarithmic mapping is defined as
% \begin{equation}
% \label{eq:log}
%     \text{log}_{\mathcal{X}_0}(\mathbf{X}_1) = \mathbf{\Gamma}.
% \end{equation}

Points on the Grassmannian are connected with smooth curves along which metrics of distances can be defined. Several such metrics exist \cite{ye2016schubert}. Perhaps the most commonly used distance metric is the \textit{geodesic} distance $d_{\mathcal{G}(p,n)}(\mathbf{X}_0, \mathbf{X}_1)$ between two points $\mathcal{X}_0, \mathcal{X}_1 \in \mathcal{G}(p,n)$, which corresponds to the distance over the geodesic $\gamma(t)$, $t \in [0,1]$, and is expressed as
\begin{equation}
\label{eq:dist}
    d_{\mathcal{G}(p,n)}(\mathbf{X}_0, \mathbf{X}_1) = {\| \mathbf{B} \|}_{2},
\end{equation}
where $\mathbf{B} = (\beta_1,\beta_2, \dots ,\beta_p)$ is the vector of principal angles obtained from the full SVD of $\mathbf{X}_0^\top\mathbf{X}_1=\mathbf{U\Sigma V}^\top$ with $\mathbf{B}=\cos^{-1}(\mathbf{\Sigma})$.

As we will see in the subsequent sections, a particularly useful way to analyze data on the Grassmannian is to embed the manifold into a Hilbert space using a valid kernel \cite{harandi2014expanding}. A Grassmannian kernel $k$ is defined as the map
\begin{equation}
\label{eq:kernel}
    k : \mathcal{G}(p,n) \times \mathcal{G}(p,n) \rightarrow \mathbb{R},
\end{equation}
where $k$ is positive semi-definite and invariant to the choice of basis.
The notion of similarity is encoded by positive semi-definite kernels on a graph and is maximized when the distance is equal to zero. Several families of Grassmannian kernels exist in the literature \cite{hamm2008extended}, the most popular being the \textit{Binet-Cauchy} and \textit{projection} kernels. 
The Binet-Cauchy kernel is used to define the Pl\"ucker embedding which maps the Grassmann manifold $\mathcal{G}(p,n)$ to the projective space $\mathbb{P}(\bigwedge^p \mathbb{R}^n)$, where the exterior product $\bigwedge ^p \mathbf{V}$ is the $k$-th product of a vector space $\mathbf{V}$. The Binet-Cauchy kernel is defined as 
\begin{subequations}
\label{eq:bc-kernel}
\begin{align}
k_{bc}(\mathbf{X}_0, \mathbf{X}_1) &= \det  (\mathbf{X}_0^\top \mathbf{X}_1)^2, \label{eq:bc-kernel_1}\\
k_{bc}(\mathbf{X}_0, \mathbf{X}_1) &= \prod_{i=1}^{p} \text{cos}^2 (\beta_i), \label{eq:bc-kernel_2}
\end{align}
\end{subequations}
where \eqref{eq:bc-kernel_2} expresses the relation between the kernel and the principal angles.
Similarly, the projection kernel is defined using the projection embedding $\Pi : \mathcal{G}(p,n) \rightarrow \mathbb{R}^{n \times n}$ given by $\Pi (\mathbf{X}) = \mathbf{X}^{T} \mathbf{X}$. Finally, the projection kernel is defined as 
\begin{subequations}
\label{eq:proj-kernel}
\begin{align}
k_{p}(\mathbf{X}_0, \mathbf{X}_1) &= {\| (\mathbf{X}_0^\top \mathbf{X}_1) \| }^2_{\text{F}},  \\
k_{p}(\mathbf{X}_0, \mathbf{X}_1) &= \sum_{i=1}^{p} \text{cos}^2 (\beta_i).
\end{align}
\end{subequations}
Throughout this work, we use the projection kernel. The interested reader is referred to \cite{dos2020grassmannian} for more information on how the Binet-Cauchy and projection kernels are constructed and applied. 

\subsubsection{Diffusion maps on the Grassmannian}
%While conventional DMAPs measures the similarity of data points in the ambient Euclidean space, GDMs consists of an initial dimension reduction step where each point in the dataset is projected onto the Grassmannian where similarity is defined by the affinity between the subspaces. One of the advantages of this step is that redundant features can be straightforwardly discarded. Once a valid Grassmannian kernel has been defined DMAPs is performed as in the following.

Consider a set of points (projected high-dimensional data) on the Grassmann manifold $\mathcal{G}(p,n)$ given by $\mathcal{G}_{N} = \{ \mathcal{X}_1,..., \mathcal{X}_N \}$ and a positive semi-definite Grassmannian kernel $k : \mathcal{G}(p,n) \times \mathcal{G}(p,n) \rightarrow \mathbb{R}$, also known as the diffusion kernel. If we consider a random walk over $\mathcal{G}_{N}$ having probability distribution $f$, $W_N = (\mathcal{G}_N, f, \mathbf{P})$, we can construct the transition probability matrix $\mathbf{P}$ as follows. First, we construct the degree matrix
\begin{equation}
    D_{ii} = \sum_{j=1}^{N} k(\mathcal{X}_i,\mathcal{X}_j),
\label{eq:diagonal}
\end{equation}
where $D_{ii}$ is a diagonal matrix $\mathbf{D} \in \mathbb{R}^{N \times N}$ and determine the stationary distribution of the random walk as
\begin{equation}
    \pi_i = \frac{D_{ii}}{\sum\limits_{k=1}^{N} D_{kk}}
\label{eq:stationary}
\end{equation}
Next, the kernel is normalized as
\begin{equation}
    \kappa_{ij} = \frac{k_{ij}}{\sqrt{D_{ii} D_{jj}}}
\label{eq:normalized}
\end{equation}
and the transition probability matrix $P_{ij}$ of the random walk over the Grassmannian is given by
\begin{equation}
    P_{ij}^t = \frac{\kappa_{ij}}{\sum\limits_{k=1}^{N} D_{ik}}.
\label{eq:trans_prob}
\end{equation}
Running the Markov chain forward in time is effectively equivalent to running a diffusion process on the manifold which allows us to reveal the geometric structure of the data on the Grassmannian. From the eigendecomposition of $\mathbf{P}^t$ we find the truncated diffusion map basis consisting of the first $q$ eigenvectors $\{ \xi_k\}_{k=1}^q$, with $\xi_k \in \mathbb{R}^N$ and corresponding eigenvalues $\{ \lambda_k\}_{k=1}^q$. Therefore, the diffusion coordinates are defined as
\begin{equation}
    \mathbf{\Theta}_j = (\theta_{j0},...,\theta_{jq}) = (\lambda_0 \xi_{j0},...,\lambda_q \xi_{jq}),
\label{eq:diff-coord}
\end{equation}
where $\xi_{jk}$ corresponds to the position $j$ of $\xi_k$. Due to the spectral decay of the eigenvalues of the sparse Markov matrix, usually a small $q$ is sufficient to capture the essential geometric structure of the dataset.

We note here that there are two essential features that distinguish the GDMaps from the conventional DMaps:
\begin{enumerate}
    \item \textbf{Data points lie on $\mathcal{G}(p,n)$:} The data on which DMaps is performed are, in fact, subspaces that compactly span the space in which the original data lie.
    \item \textbf{A Grassmannian kernel is employed:} The Grassmannian kernel is an effective means of assessing the similarity between subspaces.
\end{enumerate}
The motivation to use a subspace representation is primarily related to the difficulty in assessing similarity between very high-dimensional objects and is further elaborated in \cite{dos2020grassmannian}.

\subsection{Geometric Harmonics (GH)}
\label{gh}

Introduced by Coifman and Lafon \cite{coifman2006geometric} and based on the Nystr\"om method, GH is a method for extending an empirical function defined on a set $X$ to a set $\Bar{X}$, where $X \subset \Bar{X}$. This out-of-sample extension scheme aims to deal with the limitations of similar techniques (e.g., Kriging) related to the choice of a scale of extension.
If we assume that a real valued function $f: X \rightarrow \mathbb{R}$ is defined on $X$, GH provides a way to find an extension of $F$, say a new function $F: \Bar{X} \rightarrow \mathbb{R}$.

To begin, consider a symmetric, positive semi-definite, and bounded kernel $k: \Bar{X} \times \Bar{X} \rightarrow \mathbb{R}$, which defines a unique reproducing kernel Hilbert space $\mathcal{H}$ of functions defined on $\Bar{X}$, for which $k$ is the reproducing kernel. A typical choice is the Gaussian kernel, expressed by
\begin{equation}
    K_{ij} = k(x_i,x_j) = \exp{\Bigg(-\frac{{\| x_i - x_j   \|}^2_2}{\epsilon^2} \Bigg)},
\label{gaussian-kernel}
\end{equation}
where $\epsilon$ is a tunable length scale, $i,j=1,...,\mathcal{N}$, and ${\|\cdot\|}_2$ is the Euclidean norm. 

Given the above, it is possible to represent the function $f$ in terms of the eigenfunctions of $k$ and then extend it for out-of-sample predictions \cite{thiem2020emergent}. Given $\mathcal{N}$ realizations of the function $f(x)$, denoted $Y=\{y_i\}=\{f(x_i)\}, i=1,\dots,\mathcal{N}$, evaluated at a set of sample points $X=\{x_i\}, i=1,\dots,\mathcal{N}$, we evaluate the kernel matrix $\mathbf{K}$ with elements $K_{ij}=k(x_i,x_j)$ and perform an eigen decomposition to obtain $l$ eigenvalues, $\boldsymbol{\Lambda}=\text{diag}(\lambda_i), i=1,\dots,l$ and eigenvectors $\boldsymbol{\Psi}=[\psi_i], i=1,\dots,l$ with $l = \max_i\mid\lambda_i\ge \delta \lambda_0$. We then project the function $f$ onto the space spanned by $\psi_i, i=1,\dots,l$ 
\begin{equation}
    f \mapsto \mathbf{P}_\delta f = \sum_{j=1}^l \langle f, \psi_j \rangle_X \psi_j
\end{equation}
Practically, this is achieved by projecting the points $Y$ as $Y_\delta=\boldsymbol{\Psi}^\intercal Y$.

We then apply the Nystr\"om extension to extend the discrete eigenvectors on $X$ to eigenfunctions on $\Bar{X}$ corresponding to $\mathcal{N}_*$ out-of-sample points $\bar{x}_i, i=1,\dots,\mathcal{N}_*$. To do so, we build the extended kernel matrix as $\bar{\mathbf{K}}$ having elements $\bar{K}_{ij}=k(\bar{x}_i,x_j)$. The extended eigenvectors, or geometric harmonics, are then expressed in the matrix $\boldsymbol{\Phi}$ whose $p^{th}$ component evaluated at extension point $\bar{x}_i$ is given by
\begin{equation}
    \phi_i^{(p)} = \phi^{(p)}(\bar{x}_i) = \dfrac{1}{\lambda}_p\mathlarger{\sum}_{j=1}^{\mathcal{N}}\bar{K}_{ij} \psi_j^{(p)}.
\label{harm}
\end{equation}
Again, for practical implementation, the GH matrix is constructed simply as $\boldsymbol\Phi=\bar{\mathbf{K}}\boldsymbol\Psi \boldsymbol\Lambda^{-1}$.

Finally, we extend the function $\mathbf{P}_\delta f$ on $X$ in basis $\psi_j$ to basis $\phi_j$ on $\bar{X}$ as:
\begin{equation}
\mathbf{E}f(\bar{x})= \sum_{j=1}^l \langle f, \psi_j \rangle_X \phi_j(\bar{x})
\end{equation}
which can be performed for out-of-sample extension to a set of predicted points $\bar{Y}$ as:
\begin{equation}
    \bar{Y}=\boldsymbol\Phi Y_\delta = \boldsymbol\Phi=\bar{\mathbf{K}}\boldsymbol\Psi \boldsymbol\Lambda^{-1}\boldsymbol{\Psi}^\intercal Y
\end{equation}

% as:
% \begin{equation}
%     \Psi_j(\bar{x})=\dfrac{1}{\lambda_j}\int k
% \label{harm}
% \end{equation}

% and $\mathcal{N}_*$ points for which we want to approximate $F$ with $\ell$ geometric harmonics. We construct the kernel matrix $K_{ij}=k(x_i,x_j)$ where $K_{ij} \in \mathbb{R}^{\mathcal{N}_* \times \mathcal{N}}$, where for $x_j$ the value $f(x_j)$ is known, while $x_i$ is a new sample.  The $p$-$\text{th}$ geometric harmonic is given by
% \begin{equation}
%     \mathbf{\Phi}_i^{(p)} = \mathbf{\Phi}^{(p)}(x_i) = \mathlarger{\sum}_{j=1}^{\mathcal{N}}K_{ij} \psi_j^{(p)}\lambda^{-1}_{p},
% \label{harm}
% \end{equation}
% where $\psi^{(p)}$ is the $p$-$\text{th}$ column of the matrix $\Psi \in \mathbb{R}^{\mathcal{N} \times \ell}$ and $\lambda_p \in \Lambda = \text{diag}(\lambda_1,...,\lambda_{\ell}) \in \mathbb{R}^{\ell \times \ell}$ is its corresponding eigenvalue. Finally, the extension of function $f$ to the new function $F$ for a set of $\mathcal{N}_*$ new data points is given by 
% \begin{equation}
%     f = K\Psi\Lambda^{-1}\Psi^\intercal F
% \label{extension}
% \end{equation}

For more details and examples on GH, the reader is referred to \cite{coifman2006geometric}.

\subsection{Polynomial Chaos Expansion (PCE)}
\label{pce}

We assume a model denoted as $\mathcal{M}\left(\mathbf{X}\right)$, $\mathbf{X}$ being a $k$-variate random variable defined on the probability space $\left(\Omega, \Sigma, P\right)$ and characterized by the joint probability density function (PDF) $\varrho_{\mathbf{X}}: Z \rightarrow \mathbb{R}_{\geq 0}$, where $Z \subseteq \mathbb{R}^k$ is the image space, $\Omega$ the sample space, $\Sigma$ the set of events, and $P$ the probability measure.
Throughout this work it is assumed that $\mathbf{X}$ consists of independent random variables, however, we note that the PCE method is applicable for the case of dependent random variables as well, see e.g., \cite{feinberg2018multivariate, jakeman2019polynomial, rahman2018polynomial}.
Then, assuming that model $\mathcal{M}$ satisfies the conditions of the Doob-Dynkin lemma \cite{bobrowski2005functional}, its output $\mathcal{M}\left(\mathbf{X}\right)$ is a random variable dependent on $\mathbf{X}$.
In the following, we consider for simplicity a single model output, such that $Y\left(\omega\right) = \mathcal{M}\left(\mathbf{X}\left(\omega\right)\right) \in \mathbb{R}$, $\omega \in \Omega$. Nevertheless, the extension to multivariate outputs is straightforward, as the PCE approximation described next can be applied element-wise. Note that in the following we use the same notation for a random variable $\mathbf{X}$ and a realization $\mathbf{X}\left(\omega\right)$, however, the distinction between the two should be clear from the context.

Under the assumption of a single model output, the PCE is a spectral approximation of the form
\begin{equation}
\label{eq:spectral_approx}
\mathcal{M}(\mathbf{X}) \approx \widetilde{\mathcal{M}}(\mathbf{X}) = \sum_{s=1}^S c_s \Xi_s(\mathbf{X}),
\end{equation}
where $c_s$ are scalar coefficients and $\Xi_s$ are multivariate polynomials that are orthonormal with respect to the joint PDF $\varrho_{\mathbf{X}}$, such that
\begin{equation}
\label{eq:orthNd}
\mathbb{E}\left[\Xi_s \Xi_t\right] = \int_{Z}  \Xi_s\left(\mathbf{X}\right) \Xi_t\left(\mathbf{X}\right) \varrho_{\mathbf{X}}\left(\mathbf{X}\right) \mathrm{d}\mathbf{X} =  \delta_{st},
\end{equation}
where $\delta_{st}$ denotes the Kronecker delta.
Depending on the PDF $\varrho_{\mathbf{X}}$, the orthonormal polynomials can be chosen according to the Wiener-Askey scheme \cite{xiu2002wiener} or be numerically constructed \cite{wan2006multi, soize2004physical}.
Since $\mathbf{X}$ is assumed to consist of independent random variables $X_1, \dots, X_k$, the joint PDF is given as 
\begin{equation}
\label{eq:joint_pdf}
\varrho_{\mathbf{X}} \left(\mathbf{X}\right) = \prod_{i=1}^k \varrho_{X_i}\left(X_i\right),
\end{equation}
where $\varrho_{X_i}$ is the marginal PDF of random variable $X_i$.
Accordingly, the multivariate orthogonal polynomials are constructed as 
\begin{equation}
\label{eq:polyNd}
\Xi_s(\mathbf{X}) \equiv \Xi_\mathbf{s}(\mathbf{X}) = \prod_{i=1}^k \xi_i^{s_i} (X_i),
\end{equation}
where $\xi_i^{s_i}$ are univariate polynomials of degree $s_i \in \mathbb{Z}_{\geq 0}$ and orthonormal with respect to the univariate PDF $\varrho_{X_i}$, such that
\begin{equation}
\label{eq:orth1d}
\mathbb{E}\left[\xi_i^{s_i}\xi_i^{t_i}\right] = \int_{Z_i} \xi_i^{s_i}(X_i) \xi_i^{t_i}(X_i) \varrho_{X_i}(X_i) \, \mathrm{d}X_i = \delta_{s_i t_i}. 
\end{equation}
The multi-index $\mathbf{s} = \left(s_1, \dots, s_k\right)$ is equivalent to the multivariate polynomial degree and uniquely associated to the single index $s$ employed in Eq.~\eqref{eq:spectral_approx}, which can now be written in the equivalent form
\begin{equation}
\label{eq:spectral_approx_multi_index}
\mathcal{M}(\mathbf{X}) \approx \widetilde{\mathcal{M}}(\mathbf{X}) = \sum_{\mathbf{s} \in \Lambda} c_{\mathbf{s}} \Xi_{\mathbf{s}}(\mathbf{X}),
\end{equation}
where $\Lambda$ is a multi-index set with cardinality $\#\Lambda = S$.
The choice of the multi-index set $\Lambda$ plays a central role in the construction of the PCE, as it defines which polynomials and corresponding coefficients form the PCE.
The most common choice, as well as the one employed in this work, is that of a total-degree multi-index set, such that $\Lambda$ includes all multi-indices that satisfy $\left\| \mathbf{s} \right\|_1 \leq s_{\max}$, $s_{\max} \in \mathbb{Z}_{\geq 0}$. In that case, the size of the PCE basis is $S = \frac{\left(s_{\max} + k\right)!}{s_{\max}! k!}$, i.e., it scales polynomially with the input dimension $k$ and the maximum degree $s_{\max}$. 
For the case of high-dimensional input random variables $\mathbf{X}$, several sparse PCE algorithms have been proposed in the literature for the construction of $\Lambda$ such that the impact of the curse of dimensionality is mitigated \cite{loukrezis2020robust, blatman2011adaptive, jakeman2015enhancing, hampton2018basis, he2020adaptive, diaz2018sparse}. 

Once the multi-index set $\Lambda$ is fixed, the only thing remaining to complete the PCE is to compute the coefficients. Several approaches are suggested in the literature for computing the PCE coefficients, e.g., pseudo-spectral projection \cite{knio2006uncertainty, constantine2012sparse, conrad2013adaptive, winokur2016sparse}, interpolation \cite{buzzard2013efficient, loukrezis2019adaptive}, and, most commonly, regression \cite{blatman2011adaptive, loukrezis2020robust, hampton2018basis, diaz2018sparse, hadigol2018least, he2020adaptive, doostan2011non, tsilifis2019compressive}. The latter option is employed in this work also, such that the PCE coefficients are obtained by solving the penalized least squares problem \cite{rifkin2007notes}  
\begin{align}
\label{eq:regression}
\underset {\mathbf{c} \in \mathbb{R}^{\#\Lambda}}{\arg\min} \left\{\frac{1}{\mathcal{N}}\sum_{i=1}^{\mathcal{N}} \left( \mathcal{M}(\mathbf{X}_i) - \sum_{\mathbf{s} \in \Lambda} c_{\mathbf{s}} \Xi_{\mathbf{s}}\left(\mathbf{X}_i\right) \right)^2 + \lambda J\left(\mathbf{c}\right)\right\},
\end{align}
where $\lambda \in \mathbb{R}$ is a penalty factor, $J\left(\mathbf{c}\right)$ a penalty function acting on the vector of PCE coefficients $\mathbf{c} \in \mathbb{R}^{\#\Lambda}$, and $\mathscr{X} = \left\{\mathbf{X}_i\right\}_{i=1}^{\mathcal{N}}$ an experimental design (ED) of random variable realizations with corresponding model outputs $\mathscr{Y} = \left\{\mathbf{Y}_i\right\}_{i=1}^{\mathcal{N}}$. Common choices for the penalty function $J(\mathbf{c})$ are the $\ell_1$ and $\ell_2$ norms, in which cases problem~\eqref{eq:regression} is referred to as LASSO (least absolute shrinkage and selection operator) and ridge regression, respectively. Removing the penalty term results in an ordinary least squares (OLS) regression problem.

\section{High-Dimensional Surrogates using PCE on the Grassmannian Diffusion Manifold}
\label{method}
\begin{figure}[ht!]
\begin{center}
\includegraphics[width=0.8\textwidth]{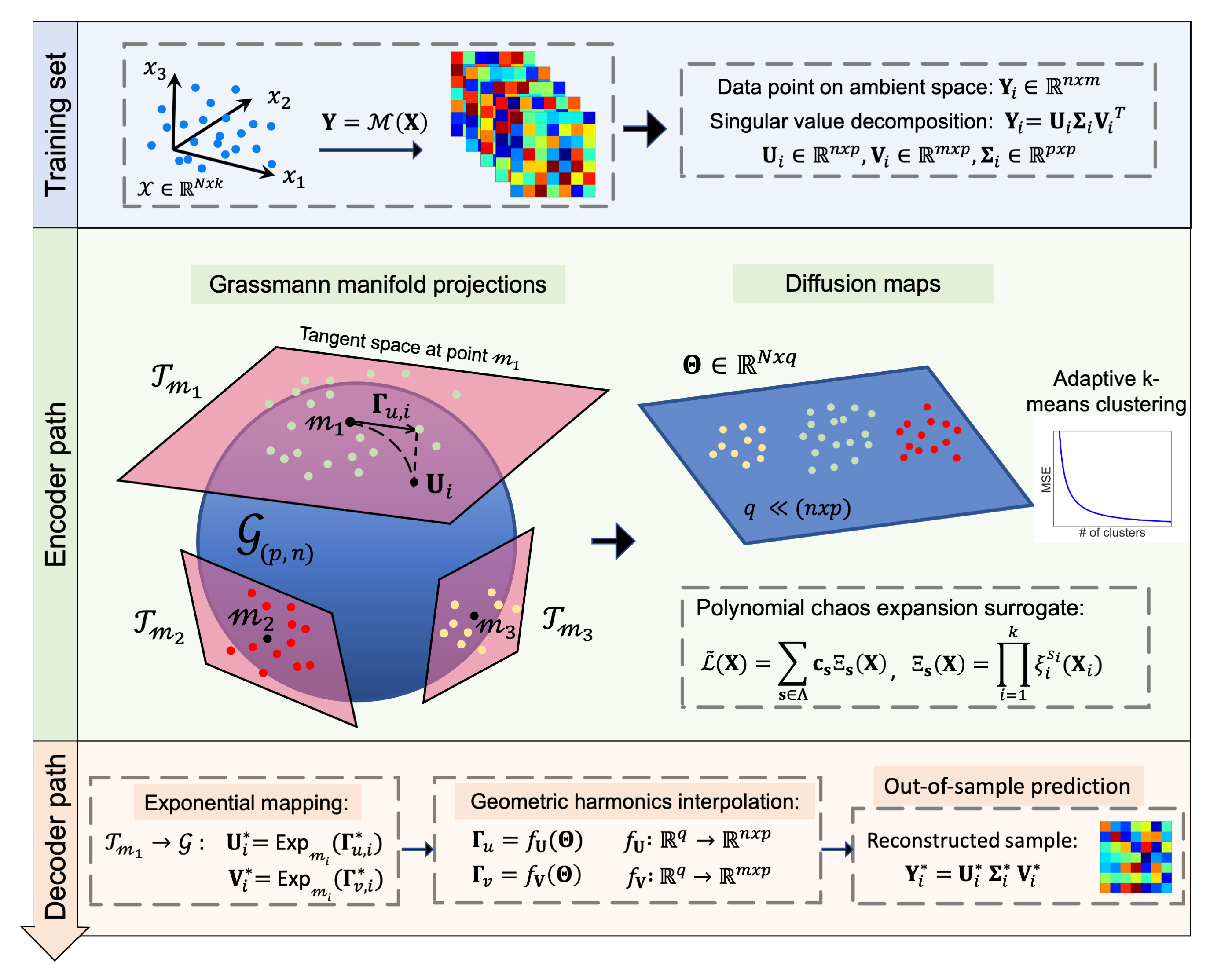}
\caption{A schematic illustrating the proposed encoder-decoder framework for constructing PCE surrogates on Grassmannian diffusion manifolds. The Grassmannian $\mathcal{G}(p,n)$ on which the set of $\{\mathbf{U}_i\}_{i=1}^\mathcal{N}$ matrices live, is depicted. A second Grassmannian $\mathcal{G}(p,m)$ exists, on which the $\{\mathbf{V}_i\}_{i=1}^\mathcal{N}$ matrices live. The diffusion manifold, represented by the diffusion coordinates $\{\mathbf{\Theta}_i\}_{i=1}^\mathcal{N}$, is also depicted. Local GH models are used to identify inverse mappings between the diffusion coordinates and points $\{\mathbf{\Gamma}_{u,i}, \mathbf{\Gamma}_{v,i}\}_{i=1}^\mathcal{N}$ that live on the various tangent spaces $\mathcal{T}_{m_i}$. Finally, exponential mappings and reverse SVD are used for sample reconstruction.}
\label{fig:schematic}
\end{center}
\end{figure}

In the next sections, the essential ingredients of the proposed method for constructing surrogate models on lower-dimensional manifolds are analytically presented.  The proposed approach is composed of two paths: 1. an encoder path where high-dimensional model data are embedded onto a low-dimensional Grassmannian diffusion manifold and a PCE surrogate is constructed to map from the input space to the low-dimensional latent space; 2. a decoder path in which predicted low-dimensional solutions in the latent space are expanded to reconstruct corresponding full, high-dimensional solutions. The approach is illustrated graphically in Figure \ref{fig:schematic}.

\subsection{Encoder Path}
The encoder path to produce an inexpensive mapping from the input space to a low-dimensional latent space representation of the high-dimensional response is detailed herein and the corresponding algorithm is provided in Algorithm \ref{algorithm-enc}.

\subsubsection{Training realizations}
\label{train-real}

% Intro- sampling etc
Consider an ED $\mathscr{X} = \{\mathbf{X}_{1},..,\mathbf{X}_{\mathcal{N}}\}$ with $\mathcal{N}$ $i.i.d.$ random samples $\mathbf{X}_{i} \in {\mathbb{R}^{k}}$ drawn from the joint PDF $\varrho_{\mathbf{X}}$. A model $\mathcal{M}: {\mathbf{X}_{i} \in \mathbb{R}^{k}} \rightarrow \mathbf{Y}_{i} \in {\mathbb{R}^{n \times m}}$ (analytical or computational) is then used to generate the corresponding model evaluations $\mathscr{Y} = \{\mathbf{Y}_{1},..,\mathbf{Y}_{\mathcal{N}}\}$. We assume that the dimensionality of the quantity of interest (QoI) $\mathbf{Y}$ is high, e.g. in the order of $\mathcal{O}(10^{4-6})$ corresponding, for example, to the number of degrees of freedom in the system, the number of time instants over which the solution is obtained, or both. We further assume that a train-test splitting procedure has already been performed for the evaluation of the surrogate model on previously unseen data. Thus, $\mathcal{N}$ represents the number of training samples which, for most real-world engineering applications, is rather small, e.g. in the order of $\mathcal{O}(10^{1-3})$. 

\subsubsection{Grassmannian Diffusion Manifold Projection}
\label{grass-proj}

For the set of $\mathcal{N}$ high-dimensional data $\mathscr{Y} = \{\mathbf{Y}_{1},..,\mathbf{Y}_{\mathcal{N}}\}$, $\mathbf{Y}_{i} \in {\mathbb{R}^{n \times m}}$, we project each data point $\mathbf{Y}_i$ onto the Grassmannian by performing a thin singular value decomposition (SVD) as
\begin{equation}
    \mathbf{Y}_i = \mathbf{U}_i \mathbf{\Sigma}_i {\mathbf{V}_i}^{\top},
\end{equation}
where the columns of the matrices (subspaces) $\mathbf{U}_i \in {\mathbb{R}^{n \times p}}$ and $\mathbf{V}_i \in {\mathbb{R}^{m \times p}}$ contain orthonormal singular vectors such that $\mathbf{U}_i^{\top}\mathbf{U}_i = \mathbf{I}_p$ and $\mathbf{V}_i^{\top}\mathbf{V}_i = \mathbf{I}_p$, and $\mathbf{\Sigma}_i \in {\mathbb{R}^{p \times p}}$ is a diagonal matrix whose non-zero elements are the singular values ordered by magnitude. Therefore, $\mathbf{U}_i, \mathbf{V}_i$ live on the Grassmannians $\mathcal{G}(p, n) = \{ \mbox{span}(\mathbf{U}) : \mathbf{U} \in \mathbb{R}^{n \times p}\}$ and $\mathcal{G}(p, m) = \{ \mbox{span}(\mathbf{V}) : \mathbf{V} \in \mathbb{R}^{m \times p}\}$, respectively. The value of dimension $p$ is either specified a priori, or computed automatically by assigning a tolerance for the SVD.

Next, for every pair $[\mathbf{U}_i, \mathbf{U}_j]$ and $[\mathbf{V}_i, \mathbf{V}_j]$ we compute the entries of $k_{ij}$ of the kernel matrices $k_{ij}(\mathbf{U})$ and $k_{ij}(\mathbf{V})$. We choose to construct either the Binet-Cauchy kernel in Eq.~\eqref{eq:bc-kernel} or the projection kernel in Eq.~\eqref{eq:proj-kernel}, thus the mappings are defined as $k_{ij}(\mathbf{U}) : \mathcal{G}(p,n) \times \mathcal{G}(p,n) \rightarrow \mathbb{R}$ and $k_{ij}(\mathbf{V}) : \mathcal{G}(p,m) \times \mathcal{G}(p,m) \rightarrow \mathbb{R}$, respectively. 
We then compute the composed kernel matrix $K(\mathbf{U},\mathbf{V})$ either by taking the sum or product of the corresponding kernels, i.e.
\begin{subequations}
\label{eq:sum-prod}
\begin{align}
    K(\mathbf{U},\mathbf{V}) &= K(\mathbf{U}) + K(\mathbf{V}), \\ K(\mathbf{U},\mathbf{V}) &= K(\mathbf{U}) \circ K(\mathbf{V}),
\end{align}
\end{subequations}
where $\circ$ denotes the Hadamard product.
The composed kernel $K(\mathbf{U},\mathbf{V})$, having components $k_{ij}$, is then used to construct the diagonal matrix $\mathbf{D} \in \mathbb{R}^{N \times N}$ in Eq.~\eqref{eq:diagonal} and then the normalized matrix $\mathbf{\kappa}$ with components $\kappa_{ij}$ in Eq.~\eqref{eq:normalized}. Next we construct the transition probability matrix $\mathbf{P}^t$ of the Markov chain over the data and we perform an eigendecomposition of $\mathbf{P}^t$ to determine the truncated diffusion map basis of $q$ eigenvectors $\{ \xi_k\}_{k=1}^q$, with $\xi_k \in \mathbb{R}^N$ and corresponding eigenvalues $\{ \lambda_k\}_{k=1}^q$. The diffusion coordinates are therefore given by $\Theta = \{\mathbf{\Theta}_{1},..,\mathbf{\Theta}_{\mathcal{N}}\}$ where $\mathbf{\Theta}_i \in \mathbb{R}^q$. Herein, we will refer to the mapping of the input parameters $\mathbf{X}$ to the diffusion coordinates $\Theta$, as $\mathcal{L}: {\mathbf{X}_{i} \in \mathbb{R}^{k}} \rightarrow \mathbf{\Theta}_{i} \in {\mathbb{R}^{q}}$.

The dimension $q$ of the diffusion coordinates $\mathbf{\Theta}_i \in \mathbb{R}^q$ (embedding) is much smaller than the dimension of the data on the ambient space $\mathbf{Y}_{i} \in {\mathbb{R}^{n \times m}}$ (i.e., $q \ll n \times m$) and therefore GDMaps allows us to achieve a significant dimension reduction.

\subsubsection{Surrogate modeling via PCE}
\label{adapt-pce}
Given a training dataset of input random variable realizations $\mathscr{X} = \{\mathbf{X}_{1},..,\mathbf{X}_{\mathcal{N}}\}$, $\mathbf{X}_i \in \mathbb{R}^k$, and corresponding solutions projected on the latent space $\Theta = \{\mathbf{\Theta}_{1},..,\mathbf{\Theta}_{\mathcal{N}}\}$, $\boldsymbol{\Theta}_i \in \mathbb{R}^q$, we construct a PCE as explained in Section \ref{pce} to approximate the true encoder $\mathcal{L}: \mathbf{X} \rightarrow \boldsymbol{\Theta}$ as
\begin{equation}
\label{eq:spectral_approx_method}
\widetilde{\mathcal{L}}(\mathbf{X}) = \sum_{\mathbf{s} \in \Lambda} \mathbf{c}_{\mathbf{s}} \Xi_{\mathbf{s}}(\mathbf{X})
\end{equation}
where $\Lambda$ is a total-degree multi-index set, $\Xi_{\mathbf{s}}$ are the multivariate orthonormal polynomials, and the PCE coefficients are now vector-valued with dimension equal to the one of the diffusion coordinates, i.e. $\mathbf{c}_{\mathbf{s}} \in \mathbb{R}^{q}$.

To assess the predictive ability of the PCE surrogate, we employ an error metric known as the \textit{generalization error} \cite{luthen2020sparse}, which is defined as
\begin{equation}
\label{eq:gen_error}
    \epsilon_{\text{gen}} = \mathbb{E}_{\mathbf{X}}\Big[\big(\mathcal{L}(\mathbf{X}) - \widetilde{\mathcal{L}}(\mathbf{X})\big)^2\Big].
\end{equation} 
We approximate $\epsilon_{\text{gen}}$ with the \textit{validation error}, which is computed on a validation dataset of $\mathcal{N}_*$ test realizations. The validation error is computed as
\begin{equation}
\label{eq:val_error}
    \epsilon_{\text{val}} = \frac{\sum_{i=1}^{\mathcal{N}_*}\big(\mathbf{\Theta}^*_i - \widetilde{\mathcal{L}}(\mathbf{X}^*_i)\big)^2}{\sum_{i=1}^{\mathcal{N}_*}\big(\mathbf{\Theta}^*_i - \bar{\mathbf{\Theta}}^*\big)^2},
\end{equation}
where $\{\mathbf{X}^*_i \in \mathbb{R}^k\}_{i=1}^{\mathcal{N}_*}$, $\{\mathbf{\Theta}^*_i \in \mathbb{R}^q\}_{i=1}^{\mathcal{N}_*}$ and $\bar{\mathbf{\Theta}}^* = \frac{1}{\mathcal{N}_*} \sum_{i=1}^{\mathcal{N}_*} \mathbf{\Theta}_i^*$ is the mean response. Accordingly, the total degree $s_{\text{max}}$ is chosen so that the validation error is minimized. In cases where a validation dataset cannot be generated due to computational constraints, alternative measures such as the \textit{k-fold cross validation} can be considered \cite{hampton2018basis, jakeman2015enhancing, luthen2020sparse}. However, such techniques introduce different computational costs, as they require the construction of multiple surrogates for different partitionings of training dataset, therefore a prior evaluation of the trade-off and respective costs is required.

{\fontsize{5}{5}\selectfont
\begin{algorithm}
\SetKwData{Left}{left}\SetKwData{This}{this}\SetKwData{Up}{up}
\SetKwFunction{Union}{Union}\SetKwFunction{FindCompress}{FindCompress}
\SetKwInOut{Input}{input}\SetKwInOut{Output}{output}
\Input{ED $\mathscr{X} = \{\mathbf{X}_{1},..,\mathbf{X}_{\mathcal{N}}\}$ where $\mathbf{X}_{i} \in {\mathbb{R}^{k}}$ and  $\mathscr{Y} = \{\mathbf{Y}_{1},..,\mathbf{Y}_{\mathcal{N}}\}$ where $\mathbf{Y}_{i} \in {\mathbb{R}^{n \times m}}$ via model $\mathcal{M}(\mathbf{X})$}
\Output{Diffusion coordinates $\Theta = \{\mathbf{\Theta}_{1},..,\mathbf{\Theta}_{\mathcal{N}}\}$  where $\mathbf{\Theta}_i \in \mathbb{R}^q$ and PCE surrogate $\widetilde{\mathcal{L}}$}
\BlankLine

\nl \For{$i\leftarrow 1$ \KwTo $\mathcal{N}$}{\nl Perform SVD
$\mathbf{Y}_i = \mathbf{U}_i \mathbf{\Sigma}_i {\mathbf{V}_i}^{\top}$ where $\mathbf{U}_i \in \mathcal{G}(p, n)$ and $\mathbf{V}_i \in \mathcal{G}(p, m)$
}
\nl Construct a Grassmannian diffusion kernel $k(\mathbf{U},\mathbf{V})$, e.g., the Binet-Cauchy via Eq.~\ref{eq:bc-kernel}

\nl Obtain the diffusion coordinates $\{\mathbf{\Theta}_i \in \mathbb{R}^q\}_{i=1}^{\mathcal{N}}$ (with Eqs.~\ref{eq:trans_prob}, \ref{eq:diff-coord}) where $q \ll n \times m$

\nl Construct PCE approximation $\widetilde{\mathcal{L}}(\mathbf{X}) = \sum_{\mathbf{s} \in \Lambda} \mathbf{c}_{\mathbf{s}} \Xi_{\mathbf{s}}(\mathbf{X})$ where $\mathbf{c}_{\mathbf{s}} \in \mathbb{R}^{q}$ can be computed via Eq.~\ref{eq:regression}

\caption{Polynomial chaos expansion on diffusion manifolds (encoder path)}
\label{algorithm-enc}
\end{algorithm}
}

\subsection{Decoder Path}

Given predictions from the PCE in the low-dimensional latent space (Grassmannian diffusion manifold), we reconstruct the approximate high-dimensional solution using the decoder path described below and detailed in Algorithm \ref{algorithm-dec}.

\subsubsection{Adaptive Clustering of Solutions on the Manifold}
\label{clustering}
An adaptive technique is proposed to cluster the diffusion coordinates $\{\mathbf{\Theta}_i\}_{i=1}^\mathcal{N}$ to iteratively identify an optimal number of clusters $\ell$, such that the distance between points belonging to one cluster is minimized both on the diffusion manifold and on the Grassmannian. We note that since the diffusion maps basis is constructed with the use of a Grassmannian kernel (affinity matrix), the notion of similarity on the Grassmannian is preserved such that points that are close on the Grassmannian are similarly close on the diffusion manifold.

To partition points at each iteration we use the $k$-means clustering algorithm. We begin with the smallest possible number of clusters, i.e., $\ell=2$. At each iteration and for each cluster $C_h$ where $h=1,..,\ell$, we compute the Karcher means $\mathbf{m}_{u,h}, \mathbf{m}_{v,h}$ from Eq.~\eqref{eq:karcher1}, of points $\{\mathbf{U}_i\}_{i=1}^{\mathcal{N}_h} \in C_h$ and $\{\mathbf{V}_i\}_{i=1}^{\mathcal{N}_h} \in C_h$, respectively, where $\mathcal{N}_h$ represents the total number of points for a given cluster. For the computation of the Karcher means and the minimization of the loss function we use stochastic gradient descent \cite{bottou2010large}. 

Having as origin the Karcher means $\mathbf{m}_{u,h}, \mathbf{m}_{v,h}$, we project the points of each cluster to the corresponding tangent spaces $\mathcal{T}_{\mathbf{m}_{u,h}}$, $\mathcal{T}_{\mathbf{m}_{v,h}}$ via the logarithmic mapping in Eq.~\eqref{eq:log}. Next we project the same points back onto the Grassmannian via the exponential mapping in Eq.~\eqref{eq:exp}. This procedure can be expressed as
\begin{subequations}
\begin{align}
  \{\mathbf{U}_i \in \mathcal{G}_{p,n}\}_{i=1}^{\mathcal{N}_h} \rightarrow \{\mathbf{\Gamma}_{u,i} \in \mathcal{T}_{\mathbf{m}_{u,h}} (\mathcal{G}_{p,n})\}_{i=1}^{\mathcal{N}_h} \rightarrow \{\mathbf{\widetilde{U}}_i \in \mathcal{G}_{p,n}\}_{i=1}^{\mathcal{N}_h} , \\
  \{\mathbf{V}_i \in \mathcal{G}_{p,n}\}_{i=1}^{\mathcal{N}_h} \rightarrow \{\mathbf{\Gamma}_{v,i} \in \mathcal{T}_{\mathbf{m}_{v,h}} (\mathcal{G}_{p,m})\}_{i=1}^{\mathcal{N}_h} \rightarrow \{\mathbf{\widetilde{V}}_i \in \mathcal{G}_{p,m}\}_{i=1}^{\mathcal{N}_h},
\end{align}
\label{proj-to-tangent-and-back}
\end{subequations}
where the subscript $h$ corresponds to a given cluster. Since the logarithmic and exponential mappings are only accurate in the local neighborhood of their origin (Karcher mean), the projection introduces error for points that deviate significantly from the Karcher mean. The point-wise error of the mapping of points $\{\mathbf{U}_i\}_{i=1}^{\mathcal{N}_h}$ and $\{\mathbf{V}_i\}_{i=1}^{\mathcal{N}_h}$ to the tangent spaces and back is computed via the mean-squared error (MSE) as
\begin{subequations}
\begin{align}
    \text{MSE}_{\mathbf{U}} = \frac{1}{\mathcal{N}_h}\sum_{i=1}^{\mathcal{N}_h}(\mathbf{U}_i - \mathbf{\widetilde{U}}_i)^2, \\
    \text{MSE}_{\mathbf{V}} = \frac{1}{\mathcal{N}_h}\sum_{i=1}^{\mathcal{N}_h}(\mathbf{V}_i - \mathbf{\widetilde{V}}_i)^2,
\end{align}
\label{mse}
\end{subequations}
respectively. Finally, the total error $\epsilon_t$ at each iteration is computed as the average of errors corresponding to each cluster. A cluster of points that are not ``close'' on the Grassmannian will result in a significant error.

We repeat this process until the error is minimized or until a specified minimum number of points has been detected in a cluster (usually $\mathcal{N}^h_{\text{min}} \simeq 5-10$). Alternatively, the process can stop when the total error is below a pre-defined threshold, e.g. $\epsilon_t < 10^{-2}$. Once the optimal number of clusters $\ell$ has been identified and the data on the manifold $\{\mathbf{\Theta}_i\}_{i=1}^\mathcal{N}$ have been appropriately partitioned in $\{C_h\}_{h=0}^\mathcal{\ell}$ clusters, out-of-sample predictions can be performed as described in the following section.

\subsubsection{Out-of-Sample Extension for High-dimensional solution prediction}
\label{geom-harm}
Consider $\mathcal{N}_*$ additional realizations of points on the diffusion manifold $\{\mathbf{\Theta}^*_i \in \mathbb{R}^q\}_{i=1}^{\mathcal{N}_*}$ have been generated using the PCE surrogate. We propose the following inverse map framework to return samples to the physically interpretable space and compute the reconstructed samples $\{\mathbf{Y}^*_i \in \mathbb{R}^{n\times m}\}_{i=1}^{\mathcal{N}_*}$.

First, we identify a mapping between training data on the diffusion manifold $\{\mathbf{\Theta}_i\}_{i=1}^\mathcal{N}$ and corresponding points on the tangent spaces of the Grassmannian $\{\mathbf{\Gamma}_{u,i}, \mathbf{\Gamma}_{v,i}\}_{i=1}^\mathcal{N}$ by constructing $2\ell$ local geometric harmonics (GH) models as described in Section \ref{gh}, two for each cluster $C_h$, $h=1,\dots, \ell$, as follows:
\begin{subequations}
\begin{align}
    \{f_{\mathbf{U}}^{C_h}\}_{h=1}^{\ell}: \mathbb{R}^{q} \rightarrow \mathbb{R}^{n \times p}, \\
    \{f_{\mathbf{V}}^{C_h}\}_{h=1}^{\ell}: \mathbb{R}^{q} \rightarrow \mathbb{R}^{m \times p},
\end{align}
\label{inverse-map}
\end{subequations}
where $f_{\mathbf{U}}^{C_h}$ and $f_{\mathbf{V}}^{C_h}$ correspond to the mapping of diffusion coordinates $\Theta^*$, to matrices $\mathbf{\Gamma}_{u}^*, \mathbf{\Gamma}_{v}^*$ respectively, for cluster $C_h$. More specifically, using the GH models, we compute the tangent space matrices as
\begin{subequations}
\begin{align}
    \mathbf{\Gamma}^*_{u,C_h,i} = f_{\mathbf{U}}^{C_h} (\mathbf{\Theta}^*_{C_h,i}), \quad i=1,\dots,\mathcal{N}_* \\
    \mathbf{\Gamma}^*_{v,C_h,i} = f_{\mathbf{V}}^{C_h} (\mathbf{\Theta}^*_{C_h,i}),  \quad i=1,\dots,\mathcal{N}_*
\end{align}
\label{inverse-tangent}
\end{subequations}
where $\mathbf{\Gamma}^*_{u,C_h,i}, \mathbf{\Gamma}^*_{v,C_h,i} $ reside on the tangent spaces  $\mathcal{T}_{C_h}\mathcal{G}(p,n)$ and $\mathcal{T}_{C_h}\mathcal{G}(p,m)$ respectively, for cluster 
$C_h$. 
% is the cluster which sample $\mathbf{\Psi}^*_i$ belongs to, and $i=1,...,\mathcal{N}_*$, where $\mathcal{N}_*$ is the total number of additional realizations.

Once the points on the tangent spaces $\{\mathbf{\Gamma}^*_{u,C_h,i} \in \mathbb{R}^{n \times p}\}_{i=1}^{\mathcal{N}_*}$ and $\{\mathbf{\Gamma}^*_{v,C_h,i} \in \mathbb{R}^{m \times p}\}_{i=1}^{\mathcal{N}_*}$ have been computed, the exponential mapping in Eq.~\eqref{eq:exp} is used to project onto the points $\{\mathbf{U}^*_{i} \in \mathbb{R}^{n \times p}\}_{i=1}^{\mathcal{N}_*}$ and $\{\mathbf{V}^*_{i} \in \mathbb{R}^{m \times p}\}_{i=1}^{\mathcal{N}_*}$ on the Grassmannians $\mathcal{G}_{p,n}$ and $\mathcal{G}_{p,m}$ (where here we drop the $C_h$ subscripts for simplicity). Finally,  we construct a global PCE surrogate to map between diffusion coordinates $\{\mathbf{\Theta}_i\}_{i=1}^\mathcal{N}$ and the diagonal singular value matrices $\{\mathbf{\Sigma}_i \in \mathbb{R}^{p \times p}\}_{i=1}^{\mathcal{N}}$, which can be used to estimate new realizations of the singular values $\{\mathbf{\Sigma}^*_i \in \mathbb{R}^{p \times p}\}_{i=1}^{\mathcal{N}^*}$ corresponding to specific out-of-sample diffusion coordinates, $\{\mathbf{\Theta}_i^*\}_{i=1}^\mathcal{N}$. 
% we construct a global PCE surrogate based on Section \ref{pce} and therefore we compute the predicted matrices of the additional realizations $\{\mathbf{\Sigma}^*_i \in \mathbb{R}^{p \times p}\}_{i=1}^{\mathcal{N}^*}$.
Reconstruction of samples is achieved by multiplying the above predicted matrices, such that
\begin{equation}
    \mathbf{Y}^*_i = \mathbf{U}^*_i \mathbf{\Sigma}^*_i \mathbf{V}^*_i,
\label{eq:recon}
\end{equation}
where $i=1,...,\mathcal{N}_*$ and $\mathbf{Y}^*_i \in \mathbb{R}^{n\times m}$.

% A schematic of the proposed framework is presented in Fig.\ \ref{fig:schematic}, illustrating the steps discussed in detail in Sections \ref{train-real}-\ref{geom-harm}. The method is also summarized in Algorithms \ref{algorithm-enc}, \ref{algorithm-dec} where the encoder and decoder path are presented separately.

{\fontsize{5}{5}\selectfont
\begin{algorithm}
\SetKwData{Left}{left}\SetKwData{This}{this}\SetKwData{Up}{up}
\SetKwFunction{Union}{Union}\SetKwFunction{FindCompress}{FindCompress}
\SetKwInOut{Input}{input}\SetKwInOut{Output}{output}
\Input{Testing samples $\mathscr{X}^* = \{\mathbf{X}_{1},..,\mathbf{X}_{\mathcal{N}_*}\}$}
\Output{Predicted solutions $\mathscr{Y}^* = \{\mathbf{Y}_{1},..,\mathbf{Y}_{\mathcal{N}}\}$ for testing samples $\mathscr{X}^*$}
\BlankLine

\nl$\ell = 2$ 

\nl \While{$\mathcal{N}^h_{\text{min}} > 5$ $\text{and}$ $\epsilon_t > 10^{-2}$}{

\nl Perform k-means clustering to identify clusters $\{C_h\}_{h=1}^{\ell}$ with $\mathcal{N}_h$ points, where $h=1,..,\ell$

\nl Compute the Karcher means $\mathbf{m}_{u,h}, \mathbf{m}_{v,h}$ from Eq.~\eqref{eq:karcher1}, of points $\{\mathbf{U}_i\}_{i=1}^{\mathcal{N}_h} \in C_h$ and $\{\mathbf{V}_i\}_{i=1}^{\mathcal{N}_h} \in C_h$

\nl Having as origin $\mathbf{m}_{u,h}, \mathbf{m}_{v,h}$, project points onto $\mathcal{T}_{\mathbf{m}_{u,h}}$, $\mathcal{T}_{\mathbf{m}_{v,h}}$ via logarithmic mapping with Eq.~\ref{eq:log}

\nl Project back to the Grassmannian via exponential mapping with Eq.~\ref{eq:exp}

\nl Compute $\text{MSE}_{\mathbf{U}}$, $\text{MSE}_{\mathbf{V}}$ with Eq.~\ref{mse}

\nl Compute total error $\epsilon_t$

\nl $\ell \leftarrow \ell+1$
}

\nl Construct $2\ell$ GH models $\{f_{\mathbf{U}}^{C_h}\}_{h=1}^{\ell}: \mathbb{R}^{q} \rightarrow \mathbb{R}^{n \times p}$ and $\{f_{\mathbf{V}}^{C_h}\}_{h=1}^{\ell}: \mathbb{R}^{q} \rightarrow \mathbb{R}^{m \times p}$ to map diff.\ coordinates $\{\mathbf{\Theta}_i\}_{i=1}^\mathcal{N}$ and points on the tangent space $\{\mathbf{\Gamma}_{u,i}, \mathbf{\Gamma}_{v,i}\}_{i=1}^\mathcal{N}$ of the Grassmannian, based on Section \ref{clustering} 

\nl Construct PCE surrogate to map $\{\mathbf{\Theta}_i\}_{i=1}^\mathcal{N}$ to singular values $\{\mathbf{\Sigma}_i \in \mathbb{R}^{p \times p}\}_{i=1}^{\mathcal{N}}$

\BlankLine
\emph{Out-of-sample prediction}

\nl Sample new points $\mathscr{X}^* = \{\mathbf{X}_{1},..,\mathbf{X}_{\mathcal{N}_*}\}$

\nl \For{$i\leftarrow 1$ \KwTo $\mathcal{N}_*$}{
\nl Compute the diffusion coordinates $\mathbf{\Theta}^*_i$  with PCE model $\widetilde{\mathcal{L}}$ 

\nl Identify the cluster $C_h$ in which point $i$ belongs to (using a Euclidean metric)

\nl Compute $\mathbf{\Gamma}^*_{u,C_h,i} = f_{\mathbf{U}}^{C_h} (\mathbf{\Theta}^*_{C_h,i})$ and $\mathbf{\Gamma}^*_{v,C_h,i} = f_{\mathbf{V}}^{C_h} (\mathbf{\Theta}^*_{C_h,i})$

\nl Perform exponential mapping to obtain $\mathbf{U}^*_{i} \in \mathbb{R}^{n \times p}$ and $\mathbf{V}^*_{i} \in \mathbb{R}^{m \times p}$

\nl Use PCE of singular values to obtain $\mathbf{\Sigma}^*_{i} \in \mathbb{R}^{p \times p}$

\nl Obtain predicted solution as $\mathbf{Y}^*_i = \mathbf{U}^*_i \mathbf{\Sigma}^*_i \mathbf{V}^*_i$
}

\nl Obtain the set of predicted solutions $\mathscr{Y}^* = \{\mathbf{Y}_{1},..,\mathbf{Y}_{\mathcal{N}_*}\}$
\caption{Polynomial chaos expansion on diffusion manifolds method (decoder path)}
\label{algorithm-dec}
\end{algorithm}
}

\section{Prediction Accuracy}

To assess the accuracy of predictions of the proposed method, we introduce three metrics. The first scalar metric is the relative $L_2$ error given by
\begin{equation}
\label{eq:L2-error}
    L_2(\mathbf{Y}_{\text{pred}}, \mathbf{Y}_{\text{ref}}) = \frac{\|\mathbf{Y}_{\text{pred}}- {\mathbf{Y}_{\text{ref}}\|}_{2}}{{\|\mathbf{Y}_{\text{ref}}\|}_{2}},
\end{equation}
where ${\|\cdot\|}_2$ denotes the standard Euclidean norm and $\mathbf{Y}_{\text{pred}}, \mathbf{Y}_{\text{ref}}$ are the prediction and reference responses respectively.
The second scalar metric we introduce is the $R^2$ score, also known as the coefficient of determination defined as
\begin{equation}
\label{eq:R2-score}
    R^2 = 1 - \frac{\mathlarger{\sum}_{i=1}^{w} \Big(\mathbf{Y}_{\text{pred}, i} - \mathbf{Y}_{\text{ref}, i}\Big)^2}{\mathlarger{\sum}_{i=1}^{w} \Big(\mathbf{Y}_{\text{ref}, i} - \overline{\mathbf{Y}}_{\text{ref}}\Big)^2},
\end{equation}
where $w$ is the total number of mesh points of the QoI and $\overline{\mathbf{Y}}_{\text{ref}}$ is the mean reference response.
The final metric we introduce is the absolute relative error which measures the error locally, in individual mesh points of the QoIs and is given by the following expression:
\begin{equation}
\label{eq:Rel-asb-error}
    \epsilon = \Bigg|\frac{\mathbf{Y}_{\text{pred}}- {\mathbf{Y}_{\text{ref}}}}{{\mathbf{Y}_{\text{ref}}}}\Bigg|.
\end{equation}
While the latter non-scalar metric is employed in the sequel to measure the error of individual realizations and provide a visual representation of them, the two scalar metrics are used to evaluate the overall accuracy of the proposed surrogate for a large number of additional realizations.

\section{Applications}
\label{examples}

\subsection{Application 1: Dielectric Cylinder in Homogeneous Electric Field}
In this first example, we consider a model problem from electromagnetic field theory, that of an infinitely long dielectric cylinder suspended in a homogeneous electric field.
Due to translational invariance along the $z$-axis the problem can be reduced to two dimensions. Thus, the computational domain is given as $\Omega = \left[-1,1\right] \times \left[-1,1\right]$ and the cylinder's domain by $D_{_\text{c}} = \left\{\mathbf{x} =\left(x,y\right) \:|\: \sqrt{x^2 + y^2} \leq r_0\right\}$, where $r_0$ is the cylinder's radius.
The dielectric material of the cylinder has the relative permittivity $\varepsilon_{\text{c}}$, while outside the cylinder's domain the relative permittivity is $\varepsilon_\text{o}$. 
The homogeneous electric field is given as $\mathbf{e}_\infty = \left(E_\infty, 0 \right)$. We further assume Dirichlet \glspl{bc} on the left and right boundaries of the rectangular domain $\Omega$ and Neumann \glspl{bc} on its top and bottom boundaries. The Dirichlet and Neumann boundaries are denoted with $\Gamma_\text{D}$ and $\Gamma_\text{N}$, respectively.
The electric potential $u(\mathbf{x})$ in $\Omega$ can be computed by solving the Laplace equation
\begin{subequations}
	\label{eq:poisson_eq_gen}
	\begin{align}
	-\nabla \cdot \left(\varepsilon \left(\mathbf{x}\right)  \nabla u\left(\mathbf{x}\right)\right)&=0, &&\mathbf{x}\in \Omega, \\
	u\left(\mathbf{x}\right)&=u^*\left(\mathbf{x}\right), &&\mathbf{x} \in \Gamma_{\text{D}}, \\
	\left(\nabla u\left(\mathbf{x}\right)\right)\cdot \mathbf{n}&=\left(\nabla u^*\left(\mathbf{x}\right)\right)\cdot \mathbf{n}, &&\mathbf{x}\in \Gamma_{\text{N}},
	\end{align}
\end{subequations}
where $\mathbf{n}$ denotes the outer normal unit vector, the permittivity $\varepsilon\left(\mathbf{x}\right)$ is given as
\begin{align}
\varepsilon(\mathbf{x})=
\begin{cases}
\varepsilon_{\text{c}}, &\mathbf{x} \in D_{\text{c}}, \\
\varepsilon_{\text{o}}, &\mathbf{x} \in \Omega \setminus D_{\text{c}},
\end{cases}
\end{align} 
and $u^*$, which is also the analytical solution to the problem, is given by
\begin{align}
\label{eq:material_prob}
u^{*}\left(\mathbf{x}\right)=-E_{\infty}x\:
\begin{cases}
1 - \frac{\varepsilon_{\text{c}} / \varepsilon_{\text{o}} - 1}{\varepsilon_{\text{c}} / \varepsilon_{\text{o}} + 1}\frac{r_0^2}{x^2 + y^2}, &\mathbf{x} \in \Omega \setminus D_{\text{c}}, \\
\frac{2}{\varepsilon_{\text{c}} / \varepsilon_{\text{o}} + 1}, &\mathbf{x} \in D_{\text{c}}.
\end{cases}
\end{align}

We consider variations in the electric potential resulting from stochasticity in two input parameters, the cylinder's radius $r_0$ and the strength of the electric field $E_{\infty}$. Information related to parameter description and values is provided in Table \ref{table:cylinder}. 

\begin{table}[ht!]
\small
\caption{Details of the input parameters of the dielectric cylinder application.}
\vspace{-7pt}
\centering
\begin{tabular}{c c l c}
\toprule
Parameters & \hspace{15pt} &  & \hspace{2pt} Uncertainty/value \\ [0.5ex]
\toprule
Cylinder radius  &   & $r_0$ & $\sim \mathcal{U}(0.20, 0.70)$\vspace{2pt}  \\ 
Strength of electric field   &   &   $E_{\infty}$ & $\sim \mathcal{U}(8, 18)$ \vspace{2pt} \\
Relative permittivity of cylinder's material   &   &  $\varepsilon_\text{c}$ &    $3$ \vspace{2pt}  \\  
Relative permittivity of surrounding space   &   &  $\varepsilon_\text{o}$ &    $1$  \vspace{1pt} \\ 
\bottomrule
\multicolumn{1}{l}{\footnotesize *$\mathcal{U}\left(a,b\right)$ denotes a uniform distribution with lower bound $a$ and upper bound $b$.} \\
\multicolumn{1}{l}{\footnotesize **All sizes are expressed in SI units.}
\end{tabular}
\label{table:cylinder}
\end{table}

\begin{figure}[ht!]
\begin{center}
\includegraphics[width=0.85\textwidth]{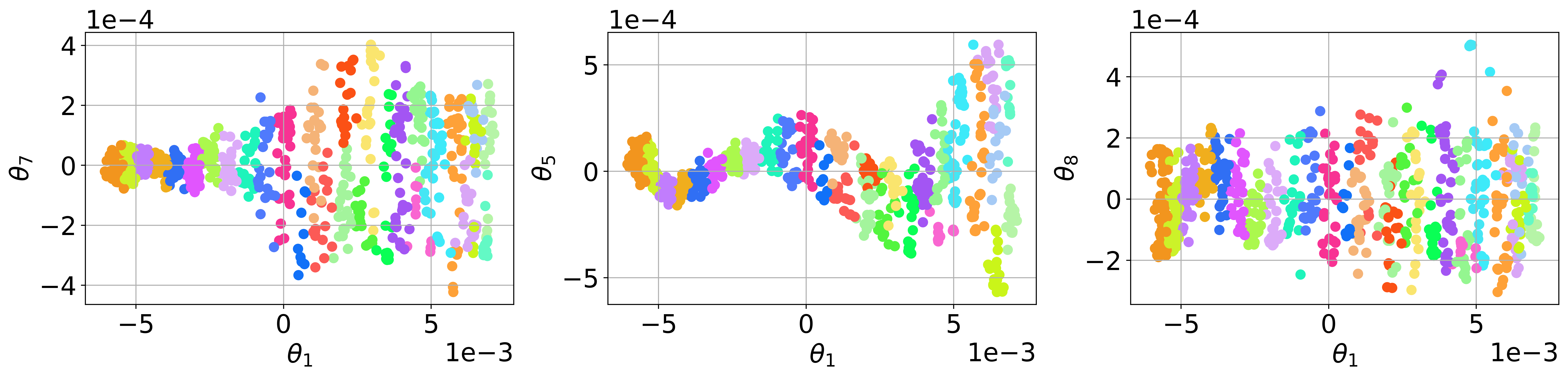}
\caption{2D plots of the diffusion coordinates $\{\theta_1, \theta_7, \theta_5, \theta_8\}$ for Grassmann manifold dimension $p=30$, $\mathcal{N}=800$ training samples, and $\ell=30$ clusters. Different colors denote the different clusters.}
\label{fig:diff-cyl}
\end{center}
\end{figure}

We generate $\mathcal{N}=\{150, 400, 800 \}$ training samples $\{\mathscr{X}_i \in \mathbb{R}^{\mathcal{N} \times 2}\}_{i=1}^{3}$, with corresponding model outputs $\{\mathscr{Y}_i \in \mathbb{R}^{\mathcal{N}\times 6400}\}_{i=1}^{3}$ where the square computational domain has been discretized in $w=(80\times80)=6400$ mesh points. GDMaps converged to a Grassmann manifold dimension of $p=30$ which results in matrices on the Grassmannian $\{\mathbf{U}_i, \mathbf{V}_i \in \mathcal{G}_{(30,80)}\}_{i=1}^{\mathcal{N}}$ for each training dataset. Based on the residuals computed by the eigendecomposition of the Markov matrix \cite{dsilva2018parsimonious}, the first $q=4$ non-trivial diffusion coordinates are considered, specifically $\{\theta_1, \theta_7, \theta_6, \theta_8\}$, $\{\theta_1, \theta_6, \theta_7, \theta_5\}$ and $\{\theta_1, \theta_7, \theta_5, \theta_8\}$, to represent the embedding structure for the three datasets with $\mathcal{N}=150$, $400$, and $800$, respectively. Therefore the method allows us to perform a dimension hyper-reduction from $\mathbb{R}^{6400}$ to $\mathbb{R}^4$, unfold the intrinsic geometric structure of the data, and reveal the essential features. A surrogate model is constructed with a maximum degree of polynomials $s_{\text{max}}=3$. The adaptive clustering algorithm converged to $\ell = \{13,25,30\}$ clusters respectively. In Figure \ref{fig:diff-cyl}, we present 2D plots of the diffusion coordinates for $\mathcal{N}=800$.

\begin{figure}[ht!]
\begin{center}
\includegraphics[width=0.85\textwidth]{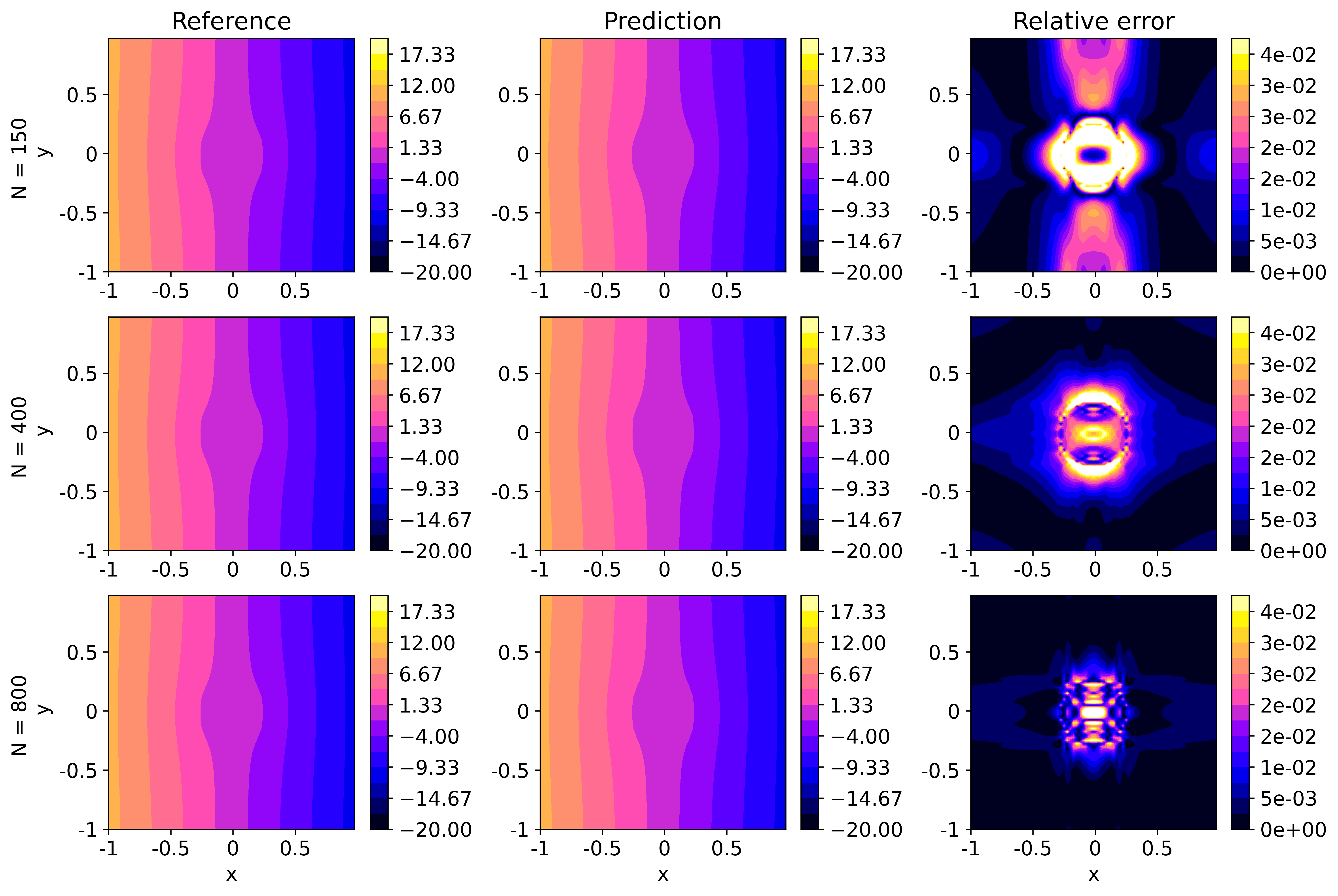}
\caption{Reference, prediction, and relative error of the electric potential field $u^{*}(x,y)$ for the random sample  $(r_0=0.273, E_{\infty}=10.523)$, $p=40$, and for $\mathcal{N}=150$ (first row), $\mathcal{N}=400$ (second row) and $\mathcal{N}=800$ (third row) training samples.}
\label{fig:comp-cyl}
\end{center}
\end{figure}

\begin{figure}[ht!]
\begin{center}
\includegraphics[width=0.85\textwidth]{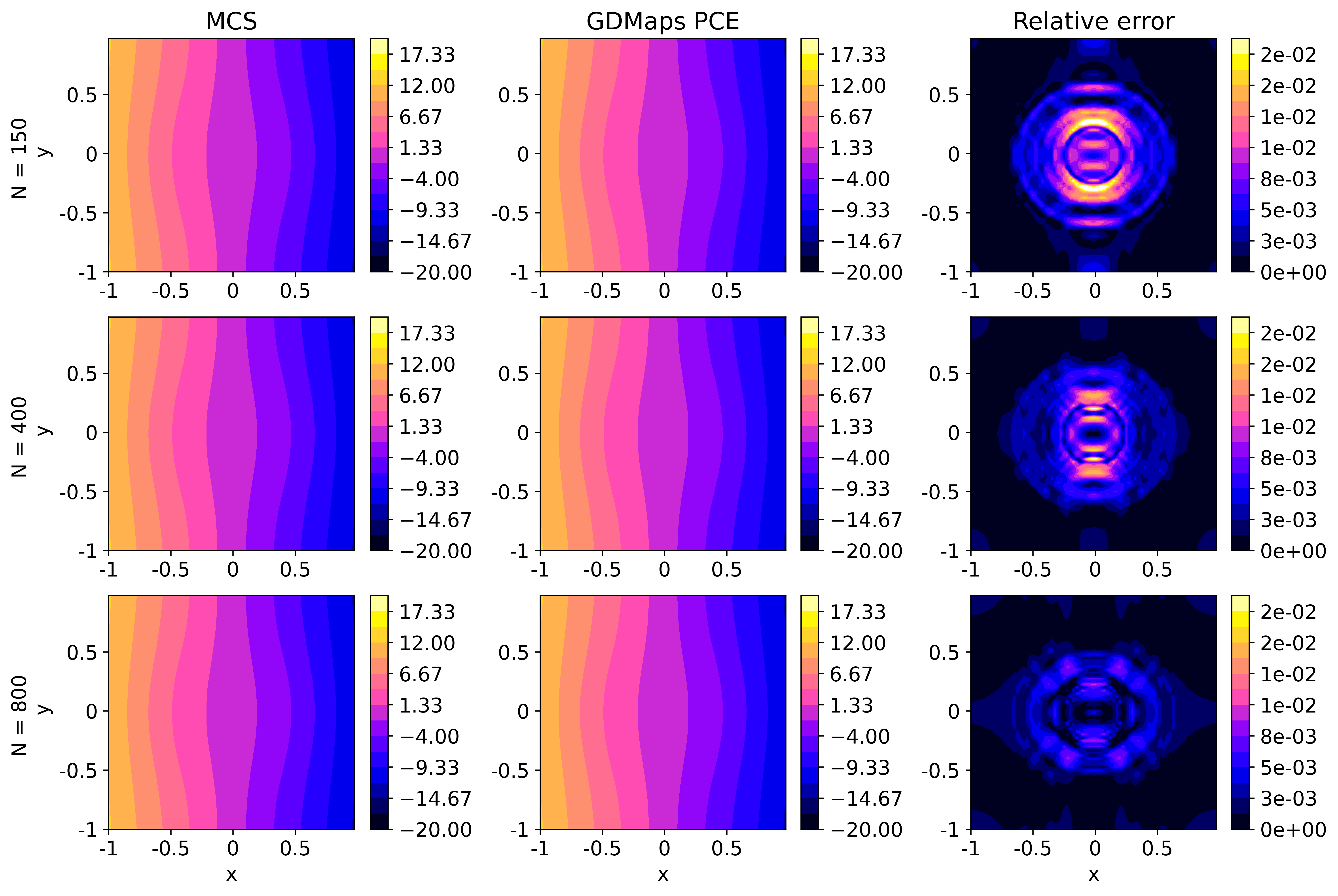}
\caption{Mean fields computed with the original model and GDMaps PCE of the electric potential field $u^{*}(x,y)$ for $\mathcal{N}=150$ (first row), $\mathcal{N}=400$ (second row) and $\mathcal{N}=800$ (third row) training samples and $10,000$ testing samples.}
\label{fig:mean-cyl}
\end{center}
\end{figure}

\begin{table}[ht!]
\small
\caption{Relative $L_2$ error and $R^2$ score for different training datasets \\ and $\mathcal{N}_* = 10,000$ testing realizations.}
\centering
\begin{tabular}{@{}ccccc@{}}
\toprule
\multirow{2}{*}{Training data} & \multicolumn{2}{c}{Relative $L_2$ error} & \multicolumn{2}{c}{$R^2$ score} \\ \cmidrule(l){2-5} 
 & Mean   & Std  & Mean  & Std  \\ \cmidrule(r){1-5}
$\mathcal{N}= 150$   & $6.704\mathrm{e}{-3}$  & $4.071\mathrm{e}{-3}$  & $9.9993\mathrm{e}{-1}$  & $8.00\mathrm{e}{-5}$  \\ $\mathcal{N}=400$  & $4.967\mathrm{e}{-3}$  & $2.644\mathrm{e}{-3}$ & $9.9996\mathrm{e}{-1}$ & $3.90\mathrm{e}{-5}$ \\
$\mathcal{N}=800$ & $4.980\mathrm{e}{-3}$   & $2.689\mathrm{e}{-3}$ & $9.9997\mathrm{e}{-1}$  & $4.00\mathrm{e}{-5}$  \\ \bottomrule
\end{tabular}
\label{table:cyl-errors}
\end{table}

\begin{figure}[ht!]
\begin{center}
\includegraphics[width=0.85\textwidth]{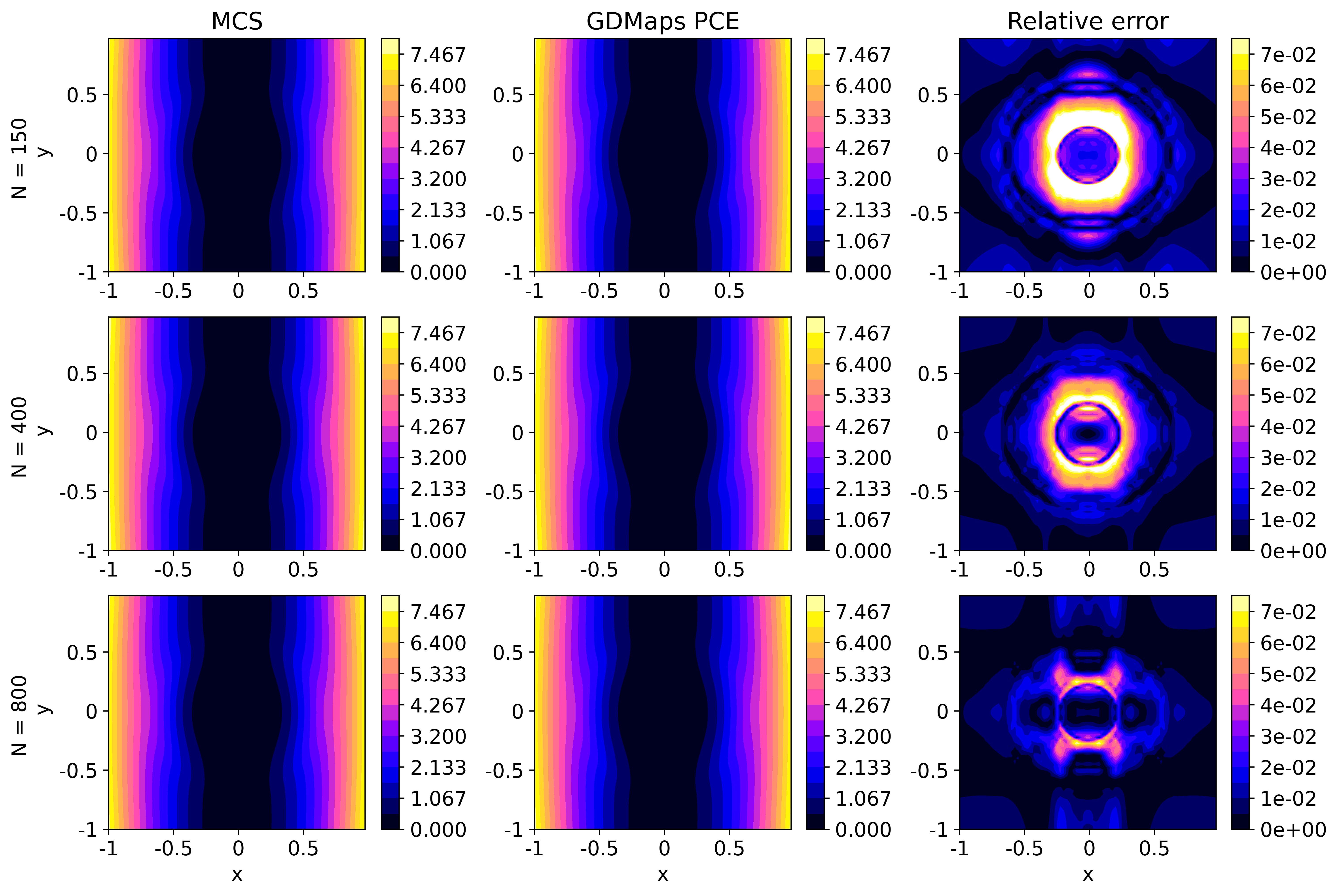}
\caption{Variance fields computed with the original model and GDMaps PCE of the electric potential field $u^{*}(x,y)$ for $\mathcal{N}=150$ (first row), $\mathcal{N}=400$ (second row) and $\mathcal{N}=800$ (third row) training samples and $10,000$ testing samples.}
\label{fig:variance-cyl}
\end{center}
\end{figure}

A comparison between the reference response and the GDMaps PCE prediction for a random sample $(r_0=0.273, E_{\infty}=10.523)$ is presented in Figure \ref{fig:comp-cyl} for all three training datasets. Overall, we observe a very good match between the reference field and surrogate predictions. The relative error is calculated based on Eq.~\eqref{eq:Rel-asb-error}, and as expected decreases as the number of training samples increases. To assess more accurately the predictive ability of the surrogate model we compute the relative $L_2$ error and coefficient of determination (or $R^2$ score) based on Eq.~\eqref{eq:L2-error} and Eq.~\eqref{eq:R2-score} respectively, for $\mathcal{N}_*=10,000$ testing realizations and we present the first two moments of the corresponding metric value distributions in Table \ref{table:cyl-errors}. Clearly, there is a significant improvement of results when we increase the number of training samples, however, we observe that the surrogate is able to perform very well in the small-data regime.

In the context of UQ, we next perform moment estimation where we compute the mean field and variance field for $\mathcal{N}_*=10,000$ with Monte Carlo simulation (MCS) on both the original model $\mathcal{M}$ and GDMaps PCE. The results for the moment and variance fields are presented in Figure \ref{fig:mean-cyl} and Figure \ref{fig:variance-cyl} respectively. In both cases, we see a very close agreement between the reference response and the surrogate prediction.

\subsection{Application 2: Lotka-Volterra Dynamical System}

In this example, we consider the classic Lotka-Volterra dynamical system \cite{mao2003asymptotic}, also known as the predator-prey equations, an example of a Kolmogorov model which describes the dynamics of a biological system in which two species interact, a predator (e.g., foxes) and a prey (e.g., rabbits). The model is a pair of non-linear ordinary differential equations (ODEs) defined as follows
\begin{equation} 
\label{lotka-volterra}
\begin{split}
    \frac{du}{dt} & = \alpha u - \beta uv, \\
    \frac{dv}{dt} & = \delta uv - \gamma v,
\end{split}
\end{equation}
where $u$ is the prey population, $v$ is the predator population and $\alpha, \beta, \gamma, \delta$ are stochastic model parameters described in Table \ref{table:lotka}. The equations have periodic solutions with $90^{\circ}$ phase difference and a linearization leads to solutions similar to those of a simple harmonic oscillator.

\begin{table}[ht!]
\small
\caption{Details of the state variables and input parameters of the Lotka-Volterra equations.}
\vspace{-7pt}
\centering
\begin{tabular}{c c l c}
\toprule
Description of variables/parameters & \hspace{15pt} &  & \hspace{2pt} Uncertainty/value \\ [0.5ex]
\toprule
Population of prey species   &  &  u  &    $u(t=0)=10$  \vspace{2pt}  \\
Population of predator species  &   & v & $v(t=0)=5$ \vspace{2pt}  \\ 
Natural growing rate of preys when no predator exists   &   &   $\alpha$ & $\sim \mathcal{U}(0.90, 1)$ \vspace{2pt} \\
Natural dying rate of preys due to predation   &   &  $\beta$ &    $\sim \mathcal{U}(0.10, 0.15)$  \vspace{2pt}  \\  
Natural dying rate of predator when no prey exists   &   &  $\gamma$ &    $1.50$  \vspace{2pt} \\ 
Reproduction rate of predators per prey eaten  &   &  $\delta$ &    $0.75$  \vspace{1pt}  \\ 
\bottomrule
\end{tabular}
\label{table:lotka}
\end{table}

For illustration, we consider two stochastic parameters $\alpha$ and $\beta$ and employ GDMaps PCE to construct a surrogate model to predict the trajectory of both predator and prey species over time.
% The initial conditions for the state variables, the probability distributions of the stochastic parameters, as well as the fixed values of the model are provided in Table \ref{table:lotka}.
We generate $\mathcal{N}=\{50, 150, 600 \}$ training samples $\{\mathscr{X}_i \in \mathbb{R}^{\mathcal{N}_i \times 2}\}_{i=1}^{3}$. For each training dataset, the system is solved using a fourth-order Runge-Kutta method with period $T=25$, discretized in $w=512$ points, thus resulting in a response matrix $\mathscr{Y} \in \mathbb{R}^{\mathcal{N}\times 1024}$ where the corresponding solutions $\{u(t), v(t)\}$ for each sample are concatenated in a single vector. Each solutions is reshaped to a square matrix $\{\mathbf{Y}_i \in \mathbb{R}^{32\times32}\}_{i=1}^{\mathcal{N}}$ and GDMaps is performed for a constant value of $p=10$, which results in matrices on the Grassmannian $\{\mathbf{U}_i, \mathbf{V}_i \in \mathcal{G}_{(10,32)}\}_{i=1}^{\mathcal{N}}$. By keeping $q=3$ parsimoniously selected diffusion coordinates we converged to the first non-trivial coordinates $\boldsymbol\Theta_i=\{\theta_1, \theta_2, \theta_5\}$, where $\{\mathbf{\Theta}_i \in \mathbb{R}^{3}\}_{i=1}^{\mathcal{N}}$. Finally, the PCE surrogate is constructed with a maximum polynomial degree $s_{\text{max}}=3$.

\begin{figure}[!ht]
\centering
\begin{minipage}{.48\textwidth}
  \centering
  \includegraphics[width=\textwidth]{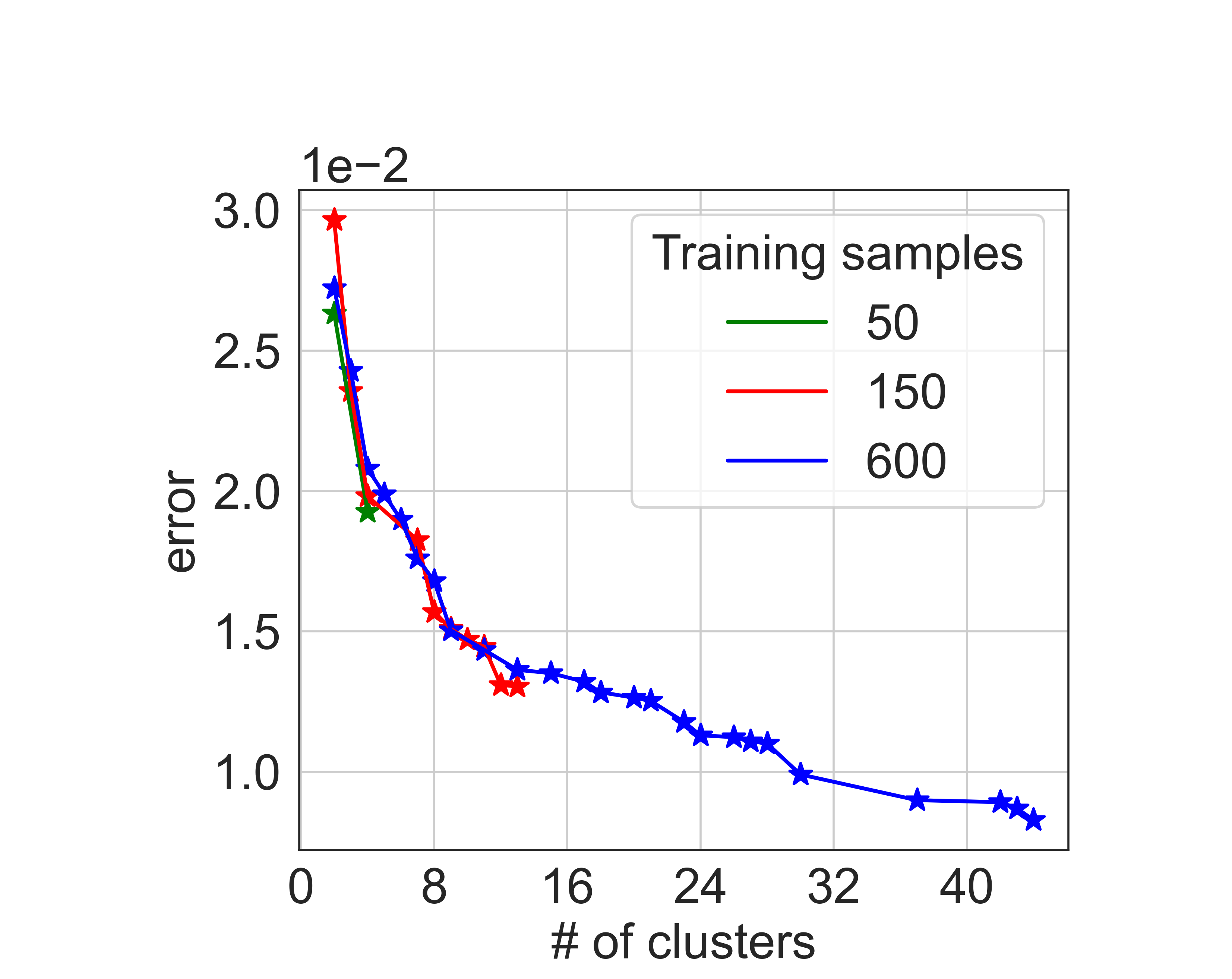}
  \captionof{figure}{Average mean square error (MSEs) associated with the projection of clusters of points $\{\mathbf{U}_i, \mathbf{V}_i \}_{i=1}^{\mathcal{N}_k}$ to the tangent space of the Grassmannian $\mathcal{G}_{(10,32)}$ and back as a function of the number of clusters for $\mathcal{N}=\{50, 150, 600 \}$.}
  \label{fig:error-lotka}
\end{minipage}%
\hspace{40pt}
\begin{minipage}{.4\textwidth}
  \centering
  \includegraphics[width=\textwidth]{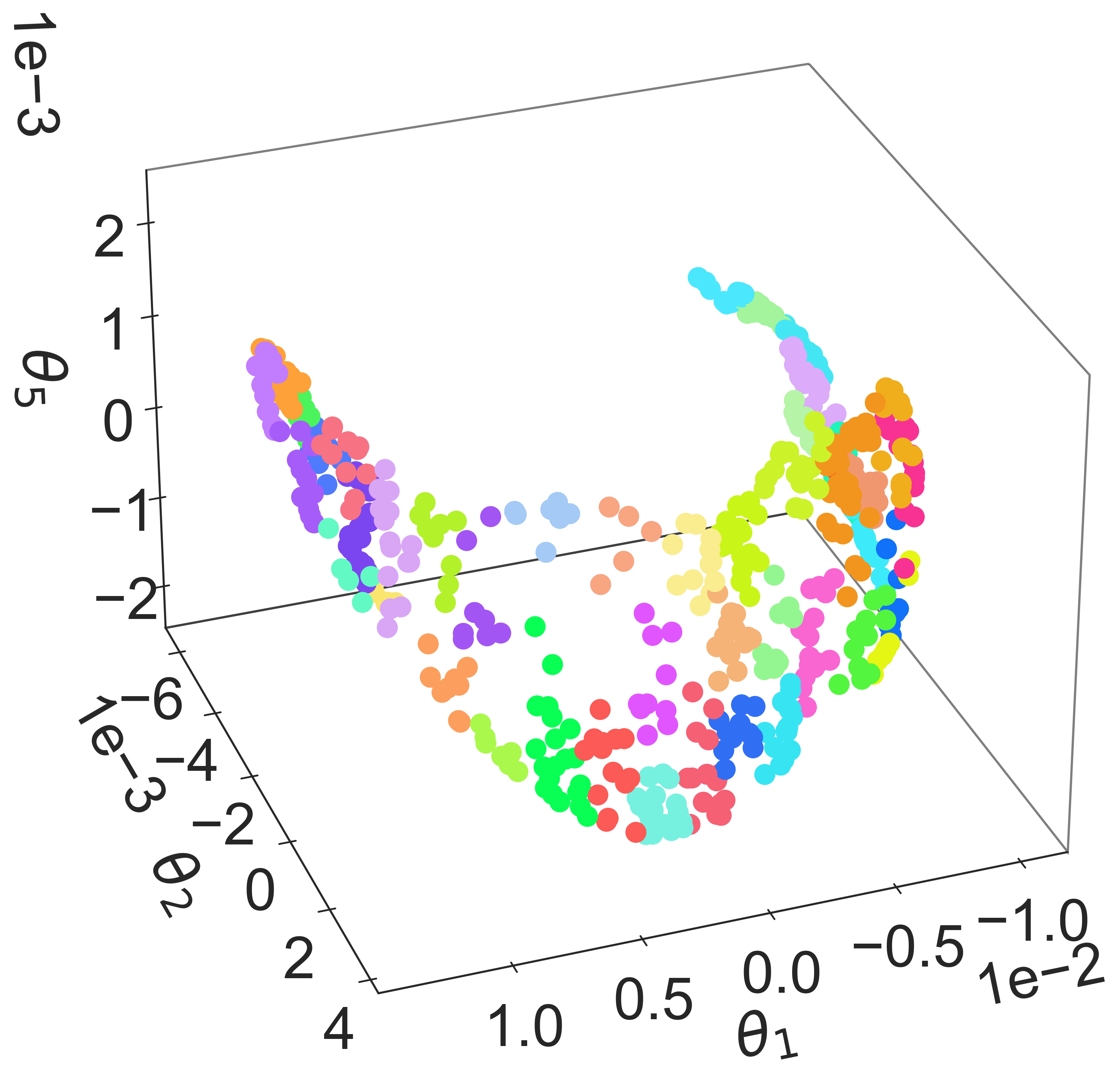}
  \captionof{figure}{Clusters of points on the Grassmannian diffusion manifold represented by the first three non-trivial diffusion coordinates $\{\theta_1, \theta_2, \theta_5\}$ for $\mathcal{N} = 600$ and $\ell = 44$.}
  \label{fig:diff-lotka}
\end{minipage}
\end{figure}

\begin{figure}[ht!]
\begin{center}
\includegraphics[width=\textwidth]{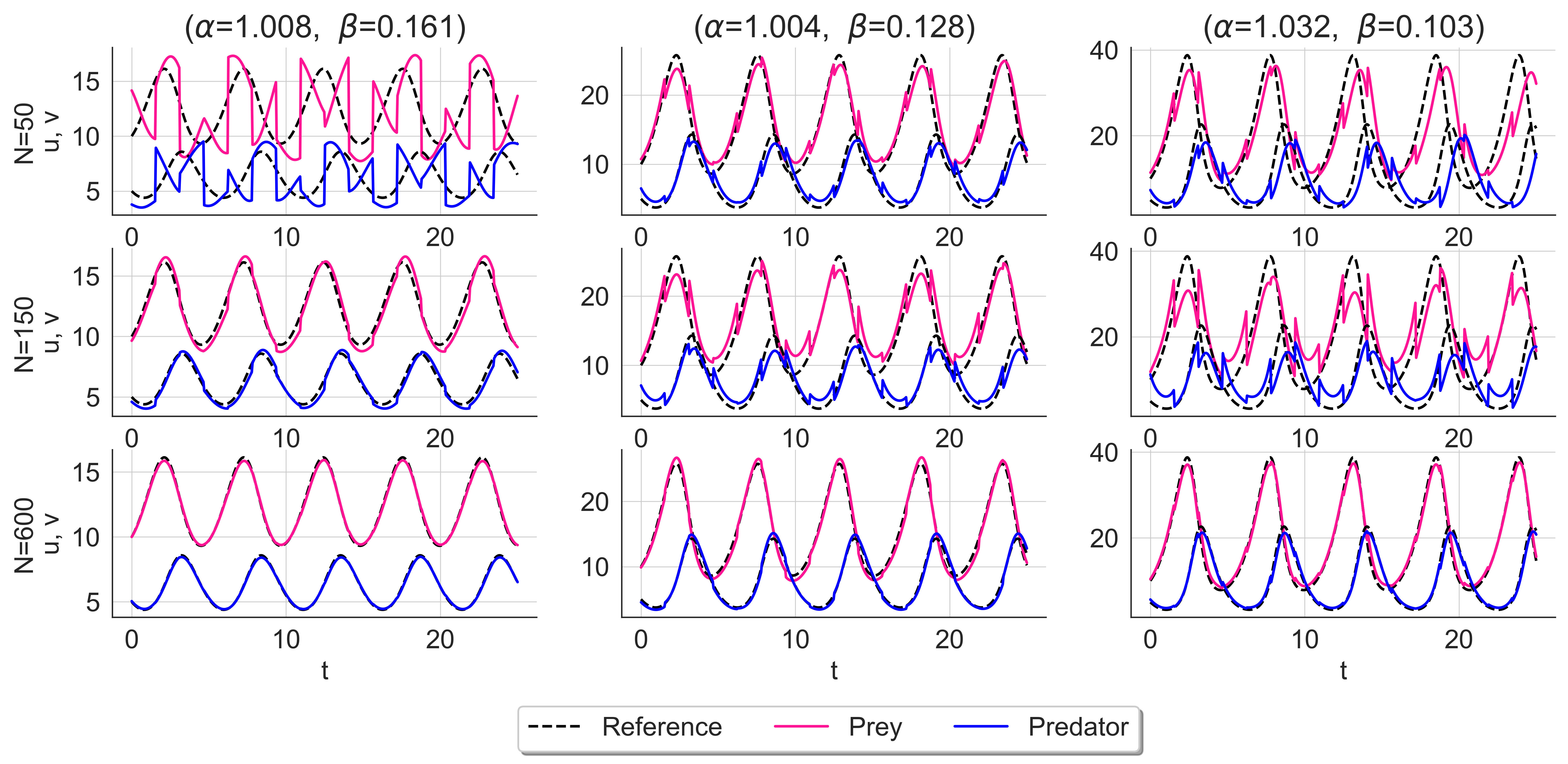}
\caption{Comparison of trajectories of reference solutions (dashed) and predictions for the prey (solid magenta) and predator (solid blue) species for $\mathcal{N}=\{50, 150, 600 \}$ (rows) and three randomly generated samples (columns).}
\label{fig:evol-lotka}
\end{center}
\end{figure}

\begin{figure}[ht!]
\begin{center}
\includegraphics[width=0.8\textwidth]{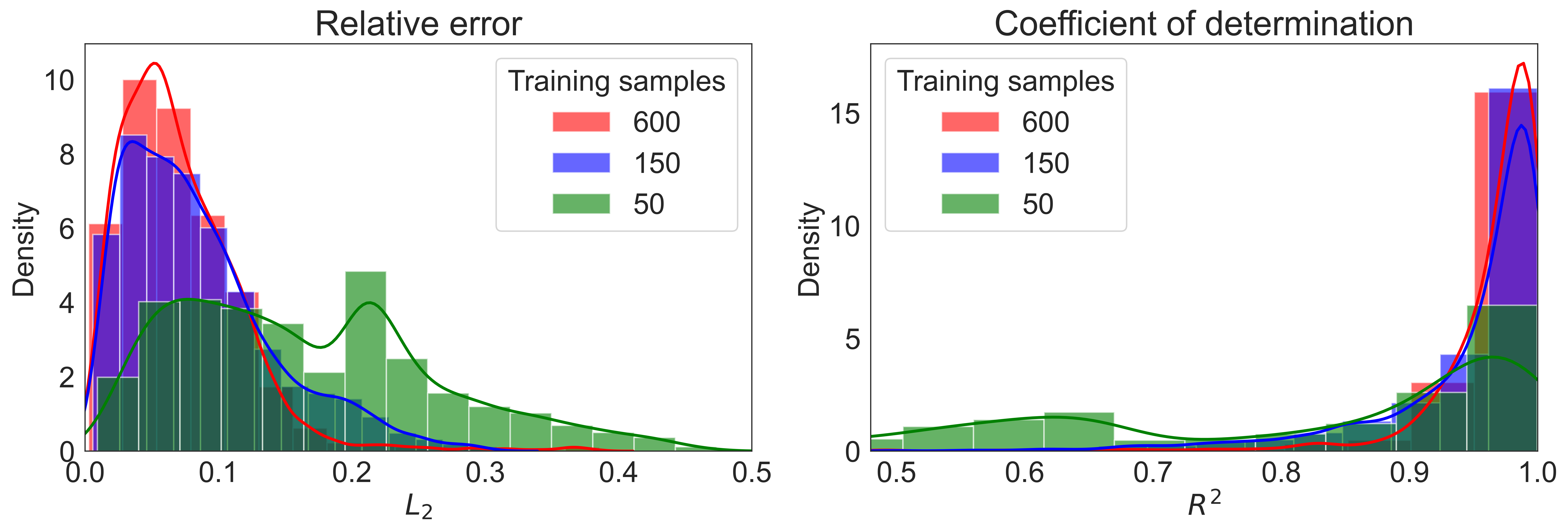}
\caption{Histograms of the relative $L_2$ error (left) and the coefficient of determination or $R^2$ score (right) of predictions and reference solutions for $\mathcal{N}=\{50, 150, 600 \}$ and $\mathcal{N}_* = 5,000$ testing realizations.}
\label{fig:hist-lotka}
\end{center}
\end{figure}

Results from the adaptive clustering algorithm for different training data set sizes are presented in Figure \ref{fig:error-lotka}. The algorithm converged to $\ell = \{4, 13, 44\}$ clusters respectively. In Figure \ref{fig:diff-lotka}, the embedding represented by the diffusion coordinates for $\mathcal{N}=600$ where $\ell=44$ is also shown. The prediction of the surrogate for the trajectories of both the prey and predator species is presented in Figure \ref{fig:evol-lotka} where we compare reference solutions and predictions for three random samples. We observe a considerable improvement of results when the number of training data increases, but nonetheless find good agreement between some points even for small training set sizes. Finally, we compare $\mathcal{N}_*=5,000$ testing realizations of the surrogate model with the corresponding reference solutions and plot the corresponding error distributions in Figure \ref{fig:hist-lotka}. We observe a rapid reduction in error (lower $L_2$, higher $R^2$) from 50 to 150 training data with continued, but less significant improvement from 150 to 600 training data. 

% that as the number of training samples increases, the distribution of the $L_2$ error is shifted to the left while the distribution of the $R^2$ score is shifted to the right, meaning that the accuracy of the surrogate model is on average improved for increasing training samples.

%A similar process is followed for a constant number of training samples $\mathcal{N}=600$ and for varying the dimension on the Grassmannian $p=\{5,10,20\}$ to determine its influence on the predictive ability of the surrogate. Similarly in Figure [REF] we compare the response for three randomly selected samples and we observe that the dimension on the Grassmannian has a significant impact on the accuracy of predictions. 

\subsection{Application 3: Advection-Diffusion-Reaction Equations}

In the third example, we consider a system of \textit{advection-diffusion-reaction} equations modeling a first-order chemical reaction between two species $A$ and $B$ that result is a formed species $C$ in some domain $\Omega$. The reaction reads
\begin{subequations}
\begin{align}
    A + B \xrightarrow{\epsilon} C, \quad \\ 
    \frac{d[C]}{dt} = K [A][B],
\end{align}
\label{eq:reaction}
\end{subequations}
where $[A],[B],[C]$ are the concentrations of the three species, $\epsilon$ is the reaction rate, and $K$ is the diffusion coefficient. According to the mass action law of chemical kinetics, the reaction rate of $C$ is proportional to the concentration of the two species $A,B$ \cite{hundsdorfer2013numerical}. 
The chemical reaction is modelled by the following set of equations:
\begin{subequations}
\begin{align}
    \dfrac{\partial [A]}{\partial t} + v \cdot \nabla [A] - \nabla \cdot (\epsilon \nabla [A]) = f_A - K[A][B], \vspace{6pt} \quad \quad \\
    \dfrac{\partial [B]}{\partial t} + v \cdot \nabla [B] - \nabla \cdot (\epsilon \nabla [B]) = f_B - K[A][B], \vspace{6pt} \quad \quad \\
    \dfrac{\partial [C]}{\partial t} + v \cdot \nabla [C] - \nabla \cdot (\epsilon \nabla [C]) = f_C + K[A][B] - K[C]
\end{align}
\label{eq:adv-system}
\end{subequations}
where species $A,B$ and $C$ diffuse throughout $\Omega$ (third terms in the left-hand side) and are advected with velocity $v$ (second terms in the left-hand side). The chemical reaction is represented in the right-hand side of Eqs.~\eqref{eq:adv-system} with source terms $f_A$, $f_B$, and $f_C$ for species $A$, $B$, and $C$ respectively. 

The chemical reaction takes place in a velocity field flowing around a cylinder and thus the above equations are coupled with the incompressible Navier-Stokes non-linear PDEs, defined as
\begin{subequations}
\begin{align}
    \rho \Big(\dfrac{\partial v}{\partial t} + v \cdot \nabla v \Big) = \nabla \cdot \sigma(v,p) + f , \qquad \vspace{6pt} \\
    \nabla \cdot v = 0 \qquad \qquad \qquad \quad
\end{align}
\label{eq:navier-stokes}
\end{subequations}
Species $A,B$ are injected into the system from two points at the top and bottom of the cylinder via the non-zero source terms $f_A, f_B$ and then advect and diffuse through the system. The third source term is set to $f_C=0$ and species $C$ is formed only as the result of the reaction of $A$ and $B$. For the numerical implementation of the above coupled system of equations we use the FEniCS package \cite{alnaes2015fenics}.

\begin{table}[ht!]
\small
\caption{Details of the state variables and input parameters of the system of \\ advection-diffusion-reaction equations.}
\vspace{-7pt}
\centering
\begin{tabular}{c c l c}
\toprule
Description of variables/parameters & \hspace{10pt} &  & \hspace{2pt} Uncertainty/value \\ [0.5ex]
\toprule
Concentration of species $A$   &  &  $[A]$  &    $[A](t=0)=0$  \vspace{2pt}  \\
Concentration of species $B$  &   & $[B]$ & $[B](t=0)=0$ \vspace{2pt}  \\ 
Concentration of species $C$  &   &   $[C]$ & $[C](t=0)=0$ \vspace{2pt} \\
Diffusion coefficient   &   &  $K$ &    $\sim \mathcal{U}(7.0, 13.0)$  \vspace{2pt}  \\  
Reaction rate   &   &  \  $\epsilon$ &    $\sim \mathcal{U}(0.005, 0.015)$ \vspace{2pt} \\ 
\bottomrule
\end{tabular}
\label{table:advection}
\end{table}

\begin{figure}[ht!]
\begin{center}
\includegraphics[width=1\textwidth]{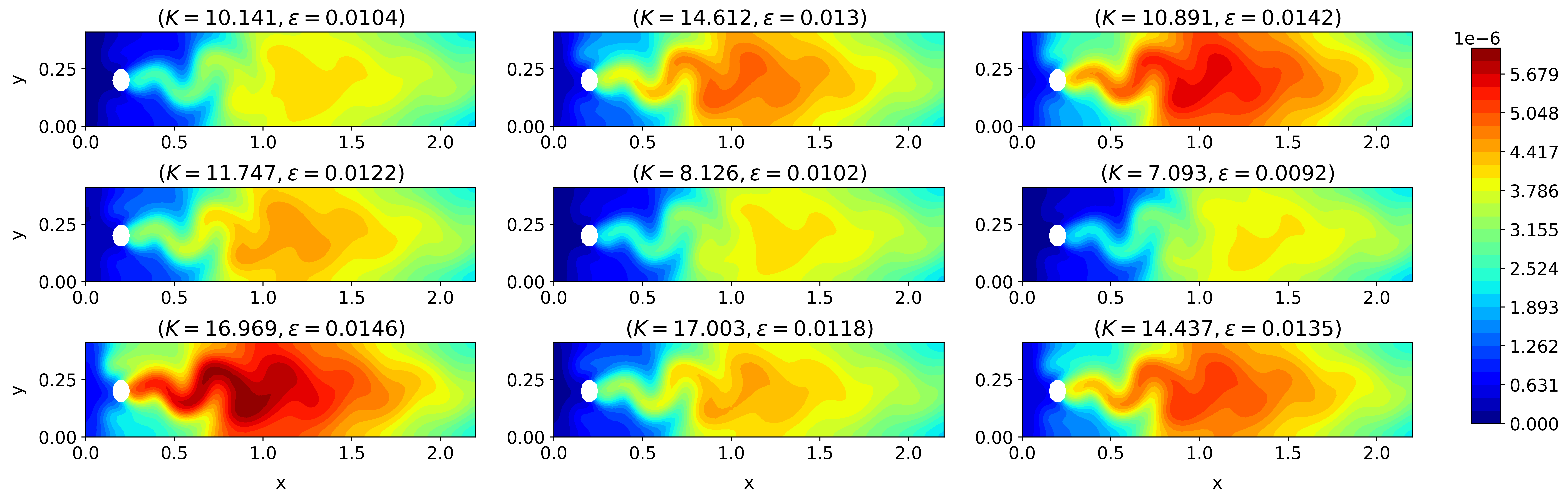}
\caption{Realizations of the stochastic field $\mathbf{Y}$, representing the concentration of species $C$ at the final time step, as a result of the chemical reaction of $A$ and $B$ with associated stochastic parameter values $\mathscr{X}$.}
\label{fig:realizations}
\end{center}
\end{figure}

We assume that stochasticity in the above system of coupled PDEs results from variations in the diffusion coefficient $K$ and the reaction rate $\epsilon$. In Table \ref{table:advection}, the initial conditions and the distributions of stochastic parameters are presented. As QoI we consider the concentration of species $C$ at the final time step, where $T=5$ is the total simulation time and $n=5000$ is the number of time steps. The simulation takes place in a rectangular domain $\Omega = \left[0,2.2\right] \times \left[0,0.41\right]$ while the cylinder is centered at $c=(0.2,0.2)$ with radius $r=0.05$. The domain is discretized with $w=2304$ mesh points. We generate $\mathcal{N}=600$ training samples $\mathbf{X}_i \in \mathbb{R}^{2}, i=1,\dots, \mathcal{N}$, and corresponding model responses $\mathbf{Y}_i \in \mathbb{R}^{2304}, i = 1,\dots,\mathcal{N}$. In Figure \ref{fig:realizations}, nine realizations of the stochastic field solution are shown. For the parametric uncertainty considered, we observe significant variations between the various profiles representing the concentration of $C$.

The solutions are reshaped to square matrices $\{\mathbf{Y}_i \in \mathbb{R}^{48\times48}\}_{i=1}^{\mathcal{N}}$ and GDMaps is performed. In this example, we aim to explore the method's predictive ability by varying the dimension of the Grassmannian on which the data are projected. We consider three values, $p=\{16,28,40\}$ which result in matrices on the Grassmannian $\{\mathbf{U}_i, \mathbf{V}_i \in \mathcal{G}_{(p,48)}\}_{i=1}^{\mathcal{N}}$. Based on the decay of eigenvalues resulting from the DMaps we retain $q=3$ of diffusion coordinates, resulting in $\{\mathbf{\Theta}_i \in \mathbb{R}^{3}\}_{i=1}^{\mathcal{N}}$. The PCE surrogate is constructed with a maximum polynomial degree, $s_{\text{max}}=3$. The adaptive clustering algorithm resulted in $\ell = \{37,31,38\}$ clusters respectively. 

\begin{figure}[ht!]
\begin{center}
\includegraphics[width=0.95\textwidth]{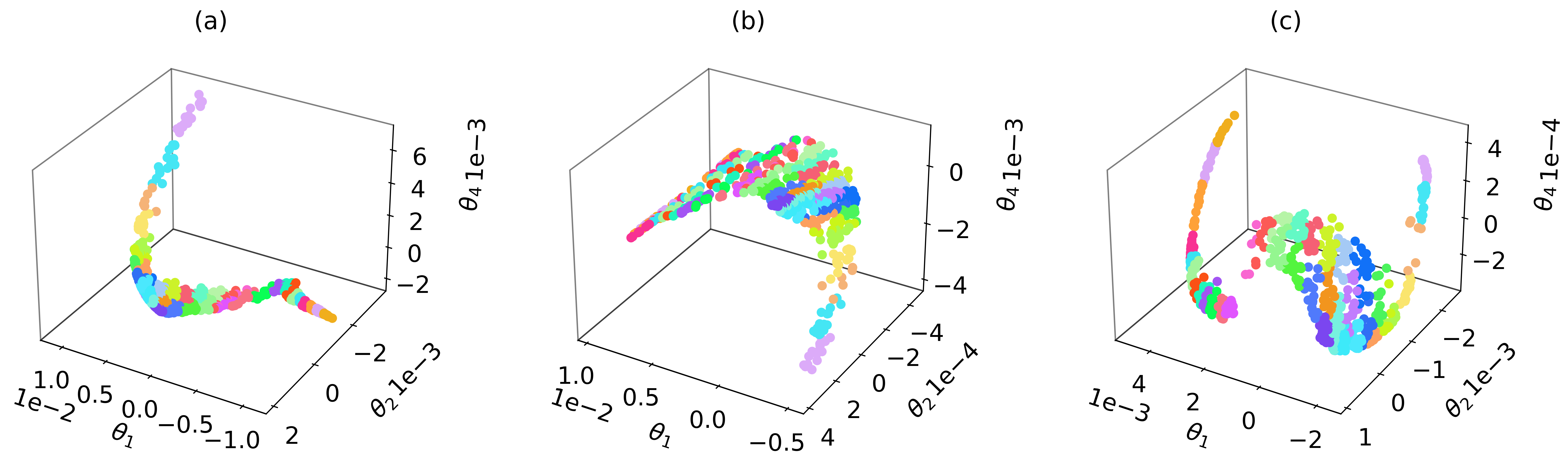}
\caption{Clustered diffusion coordinates $\{\theta_1, \theta_2, \theta_4 \}$ for Grassmann manifold dimension of (a) $p=16$, (b) $p=28$ and (c) $p=40$ for $\mathcal{N}=600$ training samples.}
\label{fig:diff-adv}
\end{center}
\end{figure}

\begin{figure}[ht!]
\begin{center}
\includegraphics[width=1\textwidth]{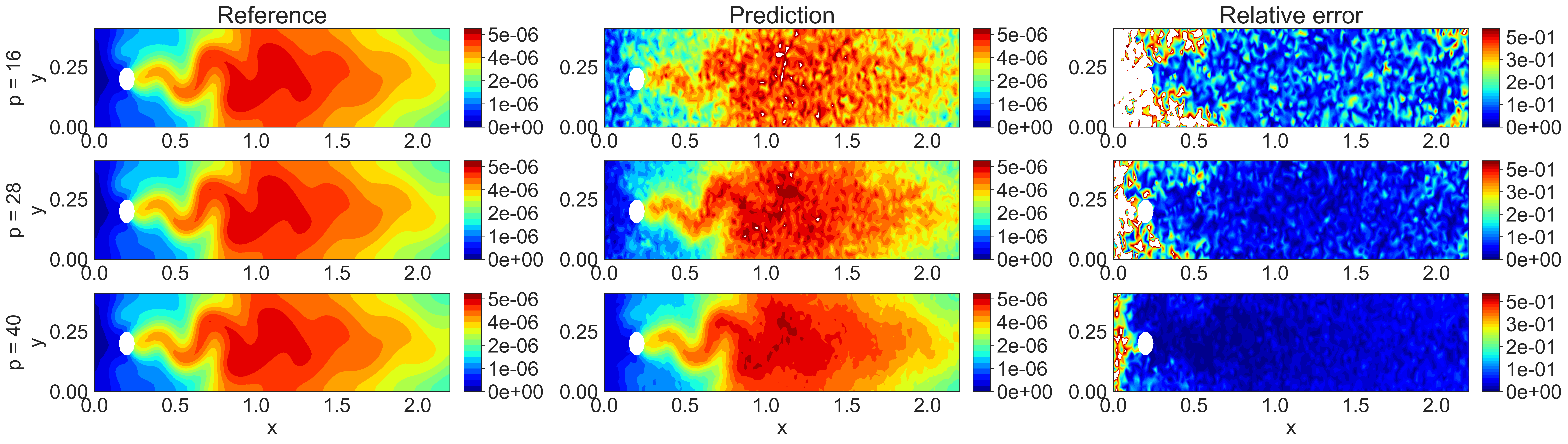}
\caption{Reference solution, surrogate prediction, and relative error of the concentration of species $C$ for parameter realizations $(K=16.138, \epsilon=0.00127)$ and Grassmann manifold dimension  $p=\{16,28,40\}$.}
\label{fig:pred-adv}
\end{center}
\end{figure}

In Figure \ref{fig:diff-adv}, we present 3D plots of the parsimoniously selected diffusion coordinates on the Grassmannian diffusion manifold for all three cases. 
% By visually observing these plots and the correlation of the coordinates, one can assess their individual significance in representing the intrinsic structure of the dataset. 
% For $p=16$, we observe that a 1D structure seems to be embedded in the 3D space, therefore coordinates $\{ \theta_2, \theta_4\}$ are not representative of new independent directions on the manifold and could be discarded. Similarly, for $p=28$ and $p=40$, coordinate $\theta_4$ appears to be non-informative and therefore the truncated DMAP basis could be sufficiently represented by the first two non-trivial coordinates $\{ \theta_1, \theta_2\}$.
In Figure \ref{fig:pred-adv}, we present the surrogate model predictions for a random sample $(K=16.138, \epsilon=0.00127)$ for $p=\{16,28,40\}$, where the reference response, the surrogate prediction, and the relative error computed with Eq.~\eqref{eq:Rel-asb-error} are shown. We observe that for a small dimension of the Grassmann manifold, the surrogate is able to capture the local intensities of the concentration of species $C$, however, with significant noise in the prediction. As we increase the dimension, we notice a significant improvement in the results and a reduction of the relative error. We observe that even though the number of diffusion coordinates remains constant in all three cases (i.e., $q=3$), the dimension of the Grassmann manifold on which the data are projected in the intermediate dimension reduction step, affects significantly the predictive ability of the surrogate. We note that the larger relative errors near the inflow boundary are caused due to the concentration (denominator) being close to zero.

\begin{figure}[ht!]
\begin{center}
\includegraphics[width=0.8\textwidth]{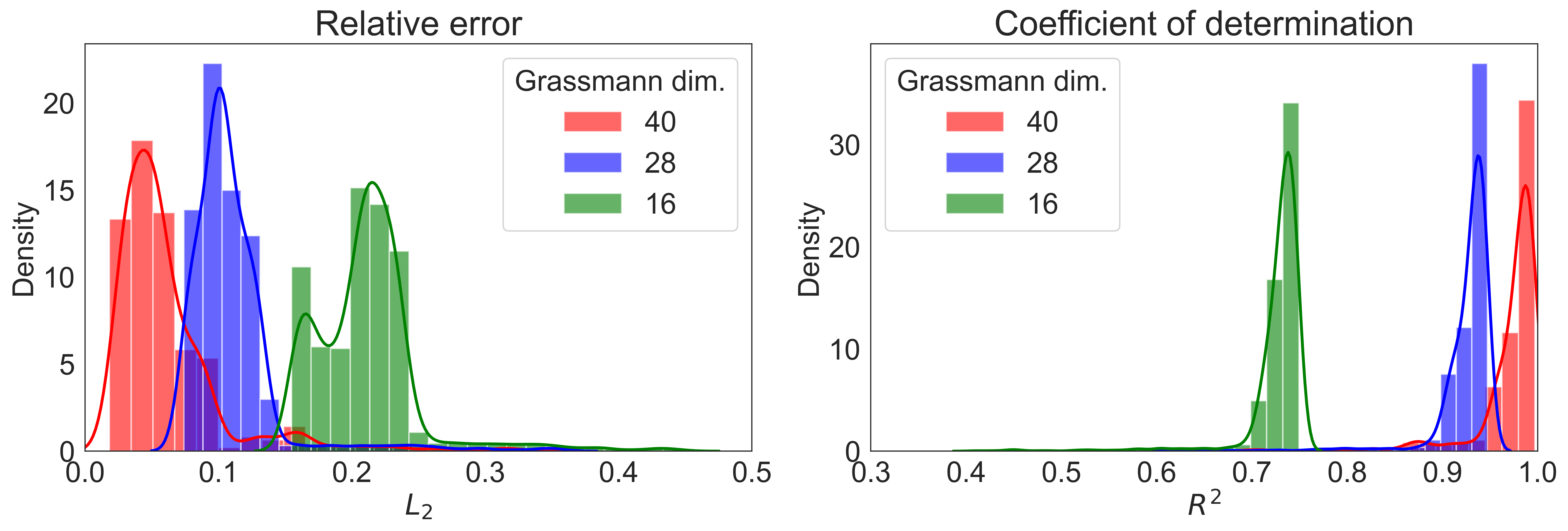}
\caption{Histograms of the relative $L_2$ error (left) and the coefficient of determination or $R^2$ score (right) of predictions and reference solutions for $\mathcal{N}_* = 1,500$ testing realizations and Grassmann manifold dimension of (a) $p=16$, (b) $p=28$ and (c) $p=40$.}
\label{fig:hist-adv}
\end{center}
\end{figure}

\begin{figure}[ht!]
\begin{center}
\includegraphics[width=1\textwidth]{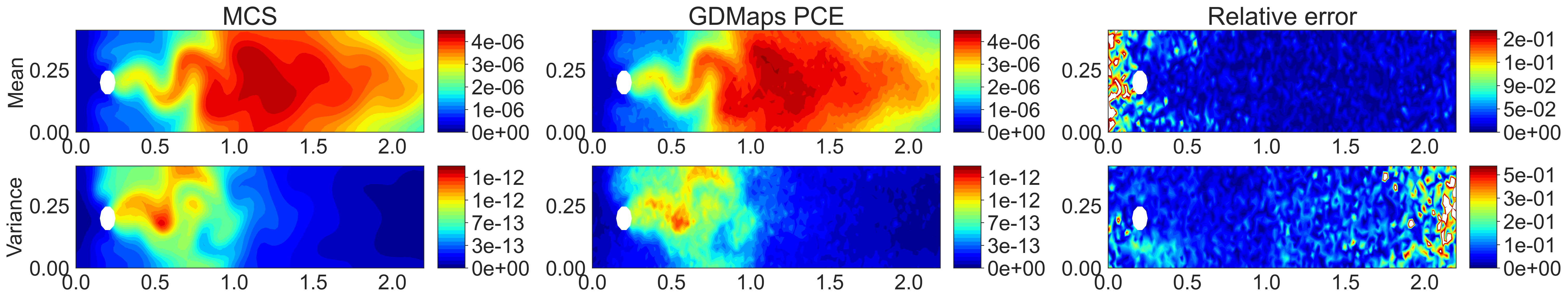}
\caption{Mean field (top row) and variance field (bottom row) for Monte Carlo simulation of $\mathcal{N}_* = 1,500$ samples, surrogate prediction and relative error for $\mathcal{N}=600$ training samples and Grassmann dimension $p=40$.}
\label{fig:mom-adv}
\end{center}
\end{figure}

To assess the overall performance of the surrogate, in Figure \ref{fig:hist-adv} we plot the distributions of the relative $L_2$ error and the $R^2$ score for $1,500$ test realizations. Clearly, the two distributions are getting closer to zero and one respectively, as the dimensionallity of the Grassmannian increases. Finally, for the same test realizations we perform moment estimation and calculate the mean and variance fields of the concentration of species $C$ via MCS with the original model and with GDMaps PCE for the same testing realizations and for $p=40$, shown in Figure \ref{fig:mom-adv}. As observed from the plots, the method is able to accurately predict the first two moments of the field.

We have demonstrated that GDMaps PCE performs very well in cases where large variability of model solutions is considered. The proposed approach results in significant cost reductions. More specifically, while a forward model evaluation requires approximately $45$ sec to complete, the proposed surrogate is able to predict model responses in an average of $0.093$ sec, i.e., $483 \times$ faster. Extrapolating these times, we see that a MC simulation of $10,000$ samples with the original model would require $\sim 5$ days of CPU time to complete, a MC simulation with GDMaps PCE would only need $\sim 15$ minutes of CPU time. The computational gains of the proposed framework become all the more prominent as the complexity, output dimensionality and, therefore, the cost of the model increases.

\section{Discussion and Conclusions}
\label{discussion}
   
This paper introduces a manifold learning-based approach for the construction of surrogate models on lower-dimensional manifolds for UQ in complex high-dimensional systems. The GDMaps PCE framework is specially designed for applications of high output dimensionality and nonlinearity. We introduced an encoder-decoder type framework in which GDMaps, a two-step dimension reduction technique for feature extraction is performed to project data onto a Grassmannian and consequently onto a diffusion manifold. Diffusion coordinates are used to represent a lower-dimensional embedding capable of capturing the salient information of the empirical dataset. A PCE surrogate is constructed on the latent space, and an adaptive clustering technique is proposed to identify regions of response similarity and consequently construct local geometric harmonics to naturally perform out-of-sample predictions. 

We explored the method’s capabilities and limitations on three applications from electromagnetic field theory, nonlinear dynamics, and chemical kinetics. Numerical results demonstrate that the proposed approach is able to accurately predict new out-of-sample solutions. The dimension of the Grassmann manifold is an important factor for the GDMaps PCE method. However, in all studied applications, we demonstrated that a very small number of coordinates $(2-4)$ representing the diffusion manifold can sufficiently capture the essential features. Furthermore, the method performed successfully under the use of small datasets and it resulted in significant reductions of the computational cost associated with the high-fidelity simulations. In the context of UQ, we have shown that GDMaps PCE provides an appropriate framework to perform statistical moment estimation in a computationally efficient manner and enables Monte Carlo simulations, which would otherwise be prohibitively expensive to compute with the original model. An interesting future direction would be to explore whether PCE coefficients computed with data on the latent space could provide useful information of QoIs in the ambient space for direct moment estimation and sensitivity analysis. 

Finally, although the proposed surrogate modeling method is ideally applicable to cases of complex models generating high-dimensional responses, its computational cost will still become intractable in cases where the input parameter space is also characterized by high-dimensionality. In such cases, methods exploiting sparse representations of PCE surrogates should be considered. These cases have not been considered. In addition, depending on the variability of the data on the diffusion manifold, an experimental design based on a standard random or quasi-random sampling technique might not be ideal. Greedy sampling techniques can be considered for such cases. These limitations form challenges to be addressed in future work.

%\section{Acknowledgements}

\bibliography{sample}

\end{document}